\documentclass[UKenglish, a4paper]{article}
\usepackage{amsmath,graphicx,commath,authblk,subcaption,float,dsfont,amssymb,booktabs,xcolor,accents,multirow,xr,hyperref}
\usepackage[utf8]{inputenc}
\usepackage[UKenglish]{babel}
\usepackage[a4paper,left=1.5cm,right=1.5cm,top=1.5cm,bottom=2cm]{geometry}
\usepackage[textwidth=4.5cm,textsize=footnotesize]{todonotes}
\usepackage[backend=biber,style=authoryear,bibstyle=authoryear,citestyle=authoryear-comp,natbib=true,maxcitenames=2,uniquelist=false,uniquename=false,sorting=ynt,datamodel=software]{biblatex}
\usepackage{software-biblatex}
\hypersetup{colorlinks=true,linkcolor=black,citecolor=black,urlcolor=cyan}
\newcommand*{\addFileDependency}[1]{
	\typeout{(#1)}
	\@addtofilelist{#1}
	\IfFileExists{#1}{}{\typeout{No file #1.}}
}\makeatother
\graphicspath{{images/}}
\DeclareMathOperator*{\argmax}{arg\,max}
\begin{document}
\author[1,$\dagger$]{N.A. Collins-Craft}
\author[2,$\star$]{I. Stefanou}
\author[1]{J. Sulem}
\author[3]{I. Einav}
\affil[1]{Laboratoire Navier, École Nationale des Ponts et Chaussées, Institut Polytechnique de Paris, Université Gustave Eiffel, CNRS, Champs-sur-Marne, France}
\affil[2]{IMSIA (UMR 9219), CNRS, EDF, CEA, ENSTA Paris, Institut Polytechnique de Paris, Palaiseau, France}
\affil[3]{Particles and Grains Laboratory, The School of Civil Engineering, The University of Sydney, Sydney, Australia}
\affil[$\dagger$]{Previously at Univ. Grenoble Alpes, Inria, CNRS, Grenoble INP, LJK, 38000, Grenoble, France}
\affil[$\star$]{Previously at GeM (Institute de Recherche en Genie Civile et Mécanique), École Centrale de Nantes, Université de Nantes, CNRS, Nantes, France}
\title{The influence of grain crushing and pore collapse on the formation of faults}
\date{}
\maketitle
\section*{Key points}\label{sec:key_points}
\begin{itemize}
	\item A new model accounts for the coevolution of the grain size distribution and porosity in crushable granular media subject to shearing.
	\item Linear stability analysis and the finite element method are used to study the localisation width of the model at depth.
	\item Orders-of-magnitude decrease in permeability is predicted to occur inside the fault core, but permeability can increase outside it.
\end{itemize}
\section*{Abstract}\label{sec:abstract}
During an earthquake, slip occurs in a localised shear zone that features a heavily granulated fault core that can be characterised as a shear band. We study the formation of this fault core in a granular rock such as sandstone by developing a model of crushable granular media within the framework of Breakage Mechanics. This model accounts for the evolution of the grain size distribution, while also accounting for the co-evolution of the solid fraction. An enrichment with the Cosserat continuum allows for the model to predict finite-width shear bands. The model is then calibrated against experimental data taken from tests on Bentheim sandstone, and a parametric study of the mechanical parameters is conducted using linear stability analysis. We find that for deeply-buried rocks the shear bands have a compactive component, and the initial value of the solid fraction does not play a strong role in the initial band thickness, but can influence the rate of delocalisation of the band. Post-localisation behaviour is studied with the finite element method, which shows the formation of zones of dilation outside the band in addition to the compaction within the band. Using a modified Kozeny--Carman permeability law, it is shown that within the band the permeability reduces by several orders of magnitude, but can increase outside the band. Our results highlight the importance of modelling grain size and solid fraction evolution as they exert a controlling influence on hydromechanical properties that play an important role in fault formation and seismic slip.
\section*{Plain language summary}\label{sec:plain_language_summary}
During an earthquake, all of the sliding motion is accommodated in a very narrow zone. Within this zone, the material has much finer grains than in the surrounding rock. To understand how this arises we develop a new model that tracks how the grains become finer, as well as how the porosity of the material reduces. The model can predict how wide the narrow zone of fine material is. We show that during the sliding process, strain localises in a narrow compacting shear band where significant grain crushing is accompanied by porosity reduction whereas porosity increase occurs outside the band. These changes in porosity and grain size strongly impact the permeability of the medium, reducing it by orders of magnitude in the compacting slip zone. This process is very important as materials through which it is difficult for fluids to flow are known to be more prone to earthquakes.
\section*{Keywords}\label{sec:keywords}
Breakage; Solid fraction; Shear bands; Cosserat continuum; Bentheim sandstone; Permeability changes
\section{Introduction}\label{sec:introduction}
Failure due to localisation into a shear band is particularly important in soils and rocks, as this failure mode is what governs the formation of a (potentially seismogenic) fault \citep{Barras2025}. While the faulting process can occur in completely dry soils and rocks in both the field \citep{Aydin1978,Aydin1983} and laboratory \citep{Ord1991}, shear band formation can also be influenced by changes to pore fluid pressure, flash melting and phase changes \citep{Rice2006,Brantut2011,Brantut2012,Veveakis2010,Veveakis2013,Rice2014a,Platt2014,Platt2015}. These processes are known to co-evolve as the fault forms and matures, with the structural consequence of the localisation of deformation into a thin band, that is typically referred to as the principal slip zone (PSZ) of the fault. The PSZ accommodates the majority of slip in a fault \citep{Chester1998,Sibson2003,Wibberley2003,Sulem2007}, and the intensity of the multiphysical processes governing the fault behaviour is controlled by the thickness of the PSZ \citep{Brantut2011,Platt2015,Sulem2016,Veveakis2013}, meaning that an accurate prediction of the evolution of this thickness is essential to accurately model fault behaviour.\\\\
Within the PSZ, the grain size distribution (GSD) features a considerably higher proportion of fine grains than the surrounding host rocks and generally follows a power-law scaling in mature faults \citep{Sammis1986,Sammis1987,Sammis1989,An1994}. The role of this microstructural property and its evolution in fault formation and behaviour has to this point been comparatively minimally studied. Aside from any effect that changes to the GSD can have on the mechanical behaviour in and of itself, any change to the GSD will also change the coupled multiphysical phenomena within the fault. As the GSD changes due to comminution, the surface area of the solid material will increase, which will in turn increase the rate of any chemical interactions that occur between the solid and the pore fluid \citep{Buscarnera2016,Stefanou2014,Zhang2018,Viswanath2019,Chen2023,Chen2024}. The evolution of the solid skeleton towards a more efficient packing will reduce the available space for the pore fluid to occupy. Simultaneously, the movement from larger to smaller pores defined by much smaller and more tightly packed grains leads to the tortuosity of the system also increasing, and as a consequence the permeability decreasing \citep{Manzocchi1998,Haines2016}. The process of shearing, and the associated frictional sliding and grain breakage will also increase the temperature, further increasing the pore pressure \citep{Rattez2017,Rattez2018a,Rattez2018b,Rattez2018c,Stathas2023} and possibly modifying the rate of any chemical reactions. All of these phenomena are tightly coupled, serving to change the effective stress experienced by the solid skeleton, which will in turn change the mechanical state by facilitating further shear deformation. This deformation leads to further changes in the GSD. This highlights the importance of being able to accurately model the evolution of the grains in the solid skeleton in order to fully capture the multiphysical behaviours that influence fault formation.\\\\
Even in cases of shallow burial depth and slow deformation (meaning thermal and chemical effects do not play any significant role), cataclastic shear band formation with significant grain breakage and porosity change is observed in the field \citep{Aydin1978,Aydin1983,Cashman2000,Fossen2007,Exner2012,Lommatzsch2015,Pizzati2020}. These bands can also be observed in laboratory experiments performed in a wide variety of apparatuses \citep{Besuelle2000,Chambon2002,Desrues2004,Sulem2006}, including in tests that are conducted without any pore fluid \citep{Rattez2022}. We may conclude that shear band formation is a pervasive feature of deforming granular rocks and soils, regardless of the presence of multiphysical couplings. This fact was demonstrated theoretically in the classic paper of \citet{Rudnicki1975} by studying the bifurcation behaviour of a general class of constitutive models appropriate for frictional materials such as soil and rocks. It has further been shown that in the classical continuum with rate-independent plasticity the favoured localisation mode is a band of infinitely small thickness \citep{Stefanou2016}. In numerical simulations such as those performed with the finite element method (FEM) this will result in a band of one element width, regardless of the mesh refinement.\\\\
The ultimate reason for this infinitely thin localisation is the absence of an internal length scale in the material model \citep{Muhlhaus1986,Muhlhaus1987}, a problem solved by specifying a model that does have such a length scale. This can be achieved by working with a model that includes viscosity, as this quantity introduces a dimension of length \citep{DeBorst2020}. However, recent results have demonstrated that for rate-dependent regularisation, strain localisation onto a plane remains theoretically possible and has been observed numerically \citep{Stathas2022}. In particular, whether the localisation onto a plane occurs in a given simulation is determined by the length of time simulated relative to the characteristic time introduced by the viscosity. We would prefer that the ability of a model to regularise localisation not be dependent on arbitrary simulation choices such as the cut-off time, but instead on the underlying physics of the system. Hence, while viscous systems may better reflect the observed rate-dependence of real materials, viscosity appears inadequate to reliably regularise the localisation of the system. Another possible method of regularisation is to include additional physics such as thermal, hydrological and chemical interactions. In this case, the internal length scale is linked to diffusion processes (see \citet{Rice2006,Brantut2017,Jacquey2021a,Heimisson2022}), particularly the outwards diffusion of pore pressure and temperature from the shear band into the bulk, as well as possibly the movement of chemical species. However, \citet{Gerolymatou2024} have recently shown that for mechanical systems coupled with physics described by an advection-diffusion equation (namely pore fluid diffusion, thermal diffusion and chemical species diffusion, as well as mechanical quantities such as the ``granular entropy'' \citep{Jiang2009}), the presence of these processes is not necessarily sufficient to guarantee regularisation. In particular, if the diffusivity depends on the strain (as permeability can), regularisation will not occur. If the advection-diffusion source term depends on the strain rate (as temperature or granular entropy do), regularisation depends on the system parameters which may change as the system evolves, making \textit{a priori} conclusions impossible. For some physical processes (\textit{e.g.} shear heating induced by friction), the parameter values will necessarily be those that do not regularise the system. As such, we can infer that multiphysical couplings are not able to \textit{a priori} regularise mechanical systems. Further, localisation in dry granular materials at room temperature is widely observed experimentally (see for instance \citet{Charalampidou2011,Karatza2018}), meaning that a purely mechanical model without multiphysical couplings is required to explain these observations, although there remains considerable interest in including hydrological, thermal and chemical effects in general models of faults \citep{Rattez2018b,Rattez2018c} to obtain a more refined description of their behaviour. The third approach to regularisation is \textit{via} a continuum with microstructure such as a nonlocal continuum \citep{Pijaudier-Cabot1988,Poh2017,Wang2018}, the second gradient continuum \citep{Zervos2001} or the Cosserat continuum \citep{Cosserat1909,Muhlhaus1987}, where the length scale is an inherent property of the system, and in geomechanics is typically related to the mean grain size of the material. As we are not aware of any theoretical or numerical demonstrations that such continua fail to regularise the system under the relevant loadings, and in light of the fact that the other regularisation approaches have been demonstrated to fail to do so (at least in certain circumstances), we conclude that the microstructured continuum approach is currently the best choice to produce a model that will regularise the system.\\\\
In granular material modelling, microstructured continuum models are most often developed in the Cosserat continuum, as there is a direct physical interpretation of the additional kinematic variable (rigid body micro-rotations) which closely resembles the observed behaviour of grains inside shear bands \citep{Ando2012,Pinzon2025}. The use of the Cosserat continuum has been further bolstered by micro-mechanical arguments in favour of a non-symmetric stress tensor \citep{Bardet2001,Papanicolopulos2011}, a characteristic feature of the continuum. Typically the internal length scale of the continuum is assumed to be a constant and unchanging multiple of the mean grain size $d_{50}$ \citep{Vardoulakis1988,Papanastasiou1989,Papanastasiou1992}. However, examination of shear bands in crushable granular media, in either field (\textit{e.g.} \citet{Rotevatn2008,Torabi2014}) or laboratory (\textit{e.g.} \citet{ElBied2002,Chambon2002}) samples demonstrates the presence of a very wide range of grain sizes that evolve under loading, meaning that this assumption is not physically realistic, and a theory describing this grain size evolution is required. This wide range of grain sizes, varying over orders of magnitude, also prevents the usage of discrete element method techniques \citep{Papachristos2023} to study the problem of particle breakage, as the computational cost of handling the extremely large number of fine particles, the extremely small time step required, and the potential contacts between particles of massively varying size is currently unreasonable. As such, a continuum mechanics theory that is capable of modelling grain breakage would seem to be the best currently available technique to address this problem.\\\\
Breakage Mechanics \citep{Einav2007,Einav2007a} is such a theory, and provides a thermodynamically consistent family of models that describe the evolution of the GSD in terms of an experimentally accessible internal variable $B$. The theory has been deployed in the development of a wide range of models, from simple models analogous to fracture mechanics theories \citep{Einav2007b,Einav2007c} to more complex models including damage \citep{Tengattini2014,Das2014,Tengattini2022a,Tengattini2022}, porosity \citep{Tengattini2016} or both \citep{Rossi2024}. The model family has also been extended to the hydrodynamic framework \citep{Alaei2022}, made rate-dependent \citep{Zhang2017,Ray2021,Ray2025}, embedded in a non-local continuum \citep{Nguyen2010}, and used to successfully model compaction band formation \citep{Das2011,Das2013a}.\\\\
In a previous work \citep{Collins-Craft2020}, we combined Breakage Mechanics with the Cosserat continuum, to develop a model describing shear bands in crushable granular media. This model features an evolving internal length scale that takes into account the entire GSD, thus rectifying the problem of an unrealistic fixed grain size that was otherwise a feature of Cosserat continuum models in geomechanics. In this paper we further refine this approach by augmenting the capacity to represent volumetric effects that occur due to comminution-driven pore collapse, dilatancy during shearing and density-dependent elasticity. In addition, we add a cohesion to the model that makes it much more suitable to modelling granular rocks by enabling an accurate description of the strength at low confining stresses. The description allows us to capture dilation without particle breakage in the case of low confining stresses, while also ensuring that under higher confining stresses the model demonstrates pore collapse accompanied by grain breakage. The inclusion of the capacity to dilate also allows us to refine the shear behaviour of the model relative to the previous one, by including a shear strength that varies with the density, as well as the Lode angle. The new model is able to recover the elastic behaviour of the model by \citet{Collins-Craft2020} as a limiting case, as well as one of the end-members of the possible plastic behaviours. Of particular importance in the context of fault mechanics, the internal variable that allows us to capture dilation also gives us sufficient information to calculate the permeability of the system, a quantity that was not accessible in the previous model, which in turn allows us to show the extent of induced permeability changes due to grain crushing, pore collapse and also dilation at the formation of a shear band. While we leave consideration of the thermal, chemical and hydraulic interactions with this mechanical model to future work, we emphasise that the model here presented is capable of predicting fault formation in realistic geological conditions and is suitable for coupling with pore fluid flow, temperature changes and chemical dissolution and precipitation, in order to a develop a comprehensive model of fault behaviour.
\section{Cosserat Breakage Mechanics formalism}\label{sec:state_characterisation}
This section discusses the set of state variables that are needed for characterising the state of a system in the Cosserat Breakage Mechanics framework, suitable for rocks and soils with distinct and crushable grains such as sands and sandstones. The section also introduces the energetic conjugates to the kinematic variables, as well as the equilibrium and boundary conditions that must be respected. All of the state variables and their dependent quantities depend on the spatial position and time, but for the sake of notational simplicity we will not write this dependence explicitly.
\subsection{Breakage state variable}\label{sec:breakage_state_variable}
\begin{figure}[H]
	\centering
	\includegraphics{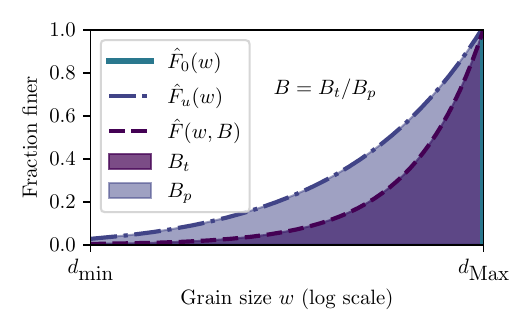}
	\caption{The ratio of two grading areas is used to define the internal breakage variable $B$, as per \citet{Einav2007a,Buscarnera2012}. $\hat{F}_{0}(w), \hat{F}_{u}(w)$ and $\hat{F}(w)=(1-B)\hat{F}_{0}(w)+B\hat{F}_{u}$ are the initial, ultimate and current cumulative GSDs respectively for a given grain size $w$. $d_{\textrm{min}}$ and $d_{\textrm{\textrm{Max}}}$ are the minimum and maximum grain sizes present in the GSDs.}
	\label{fig:breakage_plot}
\end{figure}

In order to express the evolution of the GSD under loading, we make use of the breakage state variable $B$, which is defined by the ratio of the area between the current GSD, and the initial GSD to the area between the ultimate and initial cumulative GSDs. The concept is illustrated in \figref{fig:breakage_plot}.
The maximum grain size $d_{\textrm{\textrm{Max}}}$ remains constant between the initial and ultimate distributions, and $d_{\textrm{min}}$ is the smallest grain size in the distribution. We take this to be the smallest grain that it is possible to obtain due to comminution, typically understood to be on the order of one micrometre \citep{Kendall1978}.\\\\
We use the universal initial distribution from \citet{Buscarnera2012}, by selecting a Heaviside step function about $d_{\textrm{\textrm{Max}}}$ to express the cumulative initial distribution. The corresponding probability density function of the initial distribution is given by
\begin{equation}
	\hat{p}_{0}(w)=\delta(w-d_{\textrm{Max}}), \label{eq:initial_gsd_function}
\end{equation}
where $\delta$ represents the Dirac delta function and $w$ a given grain size. We assume that the ultimate GSD follows a power law for its probability density function given by
\begin{equation}
	\hat{p}_{u}(w)=\frac{(3-\alpha)}{{d_{\textrm{Max}}}^{3-\alpha}-d_{\textrm{min}}^{3-\alpha}}w^{2-\alpha},\label{eq:ultimate_distribution}
\end{equation}
where $\alpha$ is a dimensionless constant with typical values of $2.5-2.7$.\\\\
The probability density function for the current GSD is a linear function of $B$, that expresses the current GSD in relation to the initial and ultimate GSDs:
\begin{equation}
	\hat{p}(w,B)=\hat{p}_{0}(w)(1-B)+\hat{p}_{u}(w)B. \label{eq:p(x,B)}
\end{equation}
$B$ has a corresponding thermodynamic conjugate, the breakage energy $E_{B}$, per \citet{Einav2007a,Einav2007b}.\\\\
Within the Cosserat Breakage Mechanics framework, we obtain three non-dimensional grading constants, $\theta_{\gamma}$, $\theta_{\kappa}$ and $\theta_{I}$, which express ``how far apart'' the initial and ultimate GSDs are, given by
\begin{align}
	\theta_{\gamma}&=1-\frac{3-\alpha}{5-\alpha}\left(\frac{1-(d_{\textrm{min}}/d_{\textrm{Max}})^{5-\alpha}}{1-(d_{\textrm{min}}/d_{\textrm{Max}})^{3-\alpha}}\right),\label{eq:theta_gamma_def}\\		\theta_{\kappa}&=1-\frac{3-\alpha}{7-\alpha}\left(\frac{1-(d_{\textrm{min}}/d_{\textrm{Max}})^{7-\alpha}}{1-(d_{\textrm{min}}/d_{\textrm{Max}})^{3-\alpha}}\right),\label{eq:theta_kappa_def}\\
	\theta_{I}&=1-\frac{3-\alpha}{8-\alpha}\left(\frac{1-(d_{\textrm{min}}/d_{\textrm{Max}})^{8-\alpha}}{1-(d_{\textrm{min}}/d_{\textrm{Max}})^{3-\alpha}}\right).\label{eq:theta_I_def}
\end{align}
The $\theta_{\gamma}$ term controls the energy stored due to elastic straining, the $\theta_{\kappa}$ term controls the energy stored to elastic curvatures, and the $\theta_{I}$ term controls the angular micro-rotational inertia balance. The details of the derivation of these terms are contained in \citet{Collins-Craft2020}. Physically, the higher exponents in the equations of $\theta_{\kappa}$ and $\theta_{I}$ results in their value being increasingly controlled by the largest grain size present in the GSD.\\\\
Finally, it is useful to calculate the harmonic mean grain size of the ultimate GSD
\begin{equation}
	\label{eq:ultimate_dH}
	d_{Hu}=\left[\frac{3-\alpha}{2-\alpha}\left(\frac{{d_{\textrm{Max}}}^{2-\alpha}-{d_{\textrm{min}}}^{2-\alpha}}{{d_{\textrm{Max}}}^{3-\alpha}-{d_{\textrm{min}}}^{3-\alpha}}\right)\right]^{-1},
\end{equation}
as well as the current harmonic mean grain size as it evolves with $B$:
\begin{equation}
	\label{eq:dH_with_B}
	d_{H}=\left[(1-B)d_{H0}+Bd_{Hu}\right]^{-1},
\end{equation}
where $d_{H0}$ is the harmonic mean of the initial GSD, which is simply given by
\begin{equation}
	\label{eq:initial_dH}
	d_{H0}=d_{\textrm{Max}}.
\end{equation}
\subsection{Strain and curvature rates}\label{sec:strain_and_curvature_definitions}
The following derivation adopts the small elastic strain assumption, which is appropriate for the elastic deformation range considered. This does not prevent the total deformation from being arbitrarily large, but assumes that plastic straining constantly relaxes the total straining, keeping the elastic part small \citep{Stathas2023}. We use index notation and follow the Einstein summation convention. Indices appearing after a comma represent a differentiation with respect to the spatial variable \textit{i.e.} $a_{i,j}=\frac{\partial a_{i}}{\partial x_{j}}$. The overdot represents a derivative with respect to time \textit{i.e.} $\dot{a}_{i}=\frac{\partial a_{i}}{\partial t}$, with two overdots representing the second derivative.\\\\
We define the infinitesimal (nonsymmetric) strain rate tensor by
\begin{equation}
	\dot{\gamma}_{ij}=\dot{u}_{i,j}+\epsilon_{ijk}\dot{\omega}^{c}_{k}, \label{eq:compact_strain_rate}
\end{equation}
and the infinitesimal (nonsymmetric) curvature rate tensor by
\begin{equation}
	\dot{\kappa}_{ij}=\dot{\omega}^{c}_{i,j} \label{eq:compact_curvature_rate},
\end{equation}
where $\dot{u}_{i}$ and $\dot{\omega}_{i}^{c}$ are respectively the rates of translation and rotation along and about the $x_{i}$ axes and $\epsilon_{ijk}$ is the Levi--Civita symbol. These tensors may be decomposed into symmetric and anti-symmetric parts, with the symmetric part of the strain rate tensor being the Cauchy strain rate $\dot{\varepsilon}_{ij}$. We define compression to be positive.\\\\
Both strain and curvature rate tensors can be split into trace and deviatoric parts:
\begin{align}
	\dot{\gamma}_{ij}&=\frac{1}{3}\dot{\varepsilon}_{kk}\delta_{ij}+\dot{e}_{ij},\label{eq:gamma_dot_vol_dev}\\
	\dot{\kappa}_{ij}&=\frac{1}{3}\dot{\kappa}_{kk}\delta_{ij}+\dot{z}_{ij},\label{eq:kappa_dot_vol_dev}
\end{align}
where $\delta_{ij}$ is the Kronecker $\delta$, $\dot{\varepsilon}_{kk}=\dot{\varepsilon}_{v}$ and $\dot{\kappa}_{kk}=\dot{\kappa}_{v}$.\\\\
The strain and curvature rates can be decomposed into elastic (recoverable) and plastic (nonrecoverable) parts within the framework of rate-independent plasticity:
\begin{align}
	\dot{\gamma}_{ij}&=\dot{\gamma}_{ij}^{e}+\dot{\gamma}_{ij}^{p},\label{eq:strain_decomposition}\\
	\dot{\kappa}_{ij}&=\dot{\kappa}_{ij}^{e}+\dot{\kappa}_{ij}^{p}.\label{eq:curvature_decomposition}
\end{align}
The elastic and plastic parts may also be decomposed into their respective trace and volumetric parts.
\subsection{Solid fraction state variable}\label{sec:solid_fraction_state_variable}
We adopt the solid fraction $\phi$ as a state variable, defined by
\begin{equation}
	\phi=\frac{\rho}{\rho_{s}},\label{eq:phi_def}
\end{equation}
where $\rho$ is the bulk density of the material and $\rho_{s}$ is the solid density. We choose the symbol $\phi$ for this quantity to align with the usage in the physics community. We also define the relative solid fraction $\chi$ \citep{Rubin2011}. This variable describes the solid fraction in terms of its position relative to the minimum and maximum values that are possible for a given GSD.
\begin{equation}
	\chi=\frac{\phi-\phi_{\textrm{\textrm{min}}}}{\phi_{\textrm{\textrm{Max}}}-\phi_{\textrm{min}}}, \label{eq:relative_solid_fraction}
\end{equation}
where $\phi_{\textrm{min}}$ and $\phi_{\textrm{\textrm{Max}}}$ are respectively the minimum and maximum solid fractions for a given value of $B$. We adopt the expressions of \citet{Cil2020}:
\begin{align}
	\phi_{\textrm{min}}&=1-\alpha_{\textrm{lower}}(1-B)^{l_{\phi}}\exp({-l_{\phi}B}),\label{eq:phi_min}\\
	\phi_{\textrm{\textrm{Max}}}&=1-\alpha_{\textrm{upper}}(1-B)^{u_{\phi}}\exp({-u_{\phi}B}),\label{eq:phi_max}
\end{align}
where $\alpha_{\textrm{lower}}$ and $\alpha_{\textrm{upper}}$ are terms constraining the lower and upper limiting solid densities for unbroken material and $l_{\phi}$ and $u_{\phi}$ are solid fraction bounds that can be measured experimentally.
\subsection{Stresses and couple-stresses}\label{sec:stress_and_couple-stresses}
We denote the (nonsymmetric) stress tensor as $\tau_{ij}$, and the couple-stress tensor $\mu_{ij}$, which are conjugate in power with the strain and curvature rates, respectively.\\\\
Both of these tensors can also be decomposed into trace and deviatoric parts:
\begin{align}
	\tau_{ij}&=\frac{1}{3}\tau_{kk}\delta_{ij}+s_{ij}, \label{eq:tau_vol_dev}\\
	\mu_{ij}&=\frac{1}{3}\mu_{kk}\delta_{ij}+m_{ij}, \label{eq:mu_vol_dev}
\end{align}
where $\frac{1}{3}\tau_{kk}=\frac{1}{3}\tau_{ij}\delta_{ij}$ gives the mean stress $p$. The symmetric part of the stress tensor coincides with the Cauchy stress tensor $\sigma_{ij}$.\\\\
In addition to the mean stress $p$ we have the deviatoric stress invariant $q$, and its corresponding deviatoric plastic strain rate invariant $\dot{\gamma}^{p}$. We use the formulation given in \citet{Muhlhaus1987, Vardoulakis1995, Rattez2018a} with the appropriate modifications for Breakage mechanics \citep{Collins-Craft2020}:
\begin{align}
	\dot{\gamma}^{p}&=\sqrt{g_{1}^{\star}\dot{e}_{ij}^{p}\dot{e}_{ij}^{p}+g_{2}^{\star}\dot{e}_{ij}^{p}\dot{e}_{ji}^{p}+{\ell^{e}}^{2}(g_{3}^{\star}\dot{z}_{ij}^{p}\dot{z}_{ij}^{p}+g_{4}^{\star}\dot{z}_{ij}^{p}\dot{z}_{ji}^{p})}, \label{eq:gamma_inv}\\
	q&=\sqrt{h_{1}^{\star}s_{ij}s_{ij}+h_{2}^{\star}s_{ij}s_{ji}+\frac{1}{{\ell^{e}}^{2}}(h_{3}^{\star}m_{ij}m_{ij}+h_{4}^{\star}m_{ij}m_{ji})}, \label{eq:q_inv}
\end{align}
where $g_{i}^{*}$ and $h_{i}^{*}$ are weighting factors, the values of which are given in Table~\ref{tab:weighting_factors}.
\begin{table}[H]
	\centering
	\begin{tabular}{lll}
		\hline
		& 2D model & 3D model \\\hline				
		Static model& $g_{i}^{*}=\{1/2,1/6,1/3,0\}$ & $g_{i}^{*}=\{8/15,2/15,8/15,2/15\}$ \\
		& $h_{i}^{*}=\{9/4,-3/4,3,0\}$ & $h_{i}^{*}=\{2,-1/2,2,-1/2\}$ \\ \hline
		Kinematic model & $g_{i}^{*}=\{1,-1/3,4/3,0\}$ & $g_{i}^{*}=\{8/9,-2/9,8/9,-2/9\}$ \\
		& $h_{i}^{*}=\{9/8,3/8,3/4,0\}$ & $h_{i}^{*}=\{6/5,3/10,6/5,3/10\}$ \\ \hline
	\end{tabular}
	\caption{Values of the coefficients for stress and plastic strain rate invariants in a Cosserat continuum depending on choice of dimension and static or kinematic model.}
	\label{tab:weighting_factors}
\end{table}
The expression $\ell^{e}$ is the energetic length scale defined in \citet{Collins-Craft2020} and given by
\begin{equation}
	\ell^{e}=d_{\textrm{\textrm{Max}}}\sqrt{1-\theta_{\kappa}B}. \label{eq:ell_cosserat}
\end{equation}
The Cosserat material length is no longer a constant as in \citet{Muhlhaus1987}, but depends on the entire GSD through the grading constant $\theta_{\kappa}$ and evolves as the distribution evolves with $B$.
\subsection{Equilibrium and boundary conditions}\label{sec:equilibrium_and_boundary_conditions}
We consider a thee-dimensional body $\mathcal{V}$, with a boundary $\partial\mathcal{V}$ and a moment of micro-rotational inertia $I_{ij}$. We denote the part of the boundary with Dirichlet boundary conditions as $\partial\mathcal{V}_{D}$, with prescribed velocities $\bar{\dot{u}}_{i}$ and rotation rates $\bar{\dot{\omega}}^{c}_{i}$, while the part of the boundary with Neumann boundary conditions is denoted as $\partial\mathcal{V}_{N}$, with prescribed surface tractions $\bar{\tau}_{i}$ and surface couples $\bar{\mu}_{i}$. Body forces $f_{i}$ and body couples $b_{i}$ apply everywhere in $\mathcal{V}$. The equations of motion and equilibrium equations on $\mathcal{V}$ are then \citep{Germain1973}:
\begin{equation}
	\begin{cases}
		\tau_{ij,j}-f_{i}=\rho\ddot{u}_{i},\\
		\mu_{ij,j}-\epsilon_{ijk}\tau_{jk}-b_{i}=I_{ij}\ddot{\omega}_{j}^{c},\\
		\partial\mathcal{V}_{D}\cup\partial\mathcal{V}_{N}=\partial\mathcal{V},\\
		\partial\mathcal{V}_{D}\cap\partial\mathcal{V}_{N}=\varnothing,\\
		\bar{\tau}_{i}=\tau_{ij}\bar{n}_{j}\text{ and }\bar{\mu}_{i}=\mu_{ij}\bar{n}_{j}\text{ on }\partial\mathcal{V}_{N},\\
		\dot{u}_{i}=\bar{\dot{u}}_{i}\text{ and }\dot{\omega}^{c}_{i}=\bar{\dot{\omega}}^{c}_{i}\text{ on }\partial\mathcal{V}_{D},
	\end{cases}\label{eq:momentum_balance_and_boundary_conditions}
\end{equation}
where $\bar{n}_{j}$ are the components of the normal unit vectors on $\partial\mathcal{V}$ and $\varnothing$ is the empty set.
\section{Constitutive model}\label{sec:constitutive_model}
Having detailed the state variables will be used in this work, we can now move to the specification of a particular model in the Cosserat Breakage Mechanics family. We specify only the essential elements of the new model, while leaving the detailed derivation of the model quantities to Appendix~\ref{sec:model_derivation}, and the details of its numerical implementation to the supporting~information.\\\\
We start by proposing an internal energy (per unit volume) following a limit case of the internal energy presented in \citet{Riley2025}. Unlike those authors, we ignore any consideration of pressure-dependent stiffness for the sake of simplicity, as this effect is mainly important at much lower values of confining stress than what we consider in this work. The energy potential with a suitable modification for the Cosserat continuum is:
\begin{equation}
	\hat{\mathcal{U}}(\gamma_{ij}^{e},\kappa_{ij}^{e},\rho,B)=\left(\frac{\rho}{\rho_{s}^{\star}}\right)^{n}\left(\frac{1}{2}(1-\theta_{\gamma}B)C_{ijkl}^{e}\gamma_{ij}^{e}\gamma_{kl}^{e}+\frac{1}{2}(1-\theta_{\kappa}B){d_{\textrm{Max}}}^{2}D_{ijkl}^{e}\kappa_{ij}^{e}\kappa_{kl}^{e}\right),\label{eq:u_function_with_rho_star}
\end{equation}
where $\rho_{s}^{\star}$ is the unstressed solid density that is a fixed material constant, $n$ is the degree of nonlinearity in the density, and
\begin{align}
	C_{ijkl}^{e}&=\left(\bar{K}-\frac{\bar{2G}}{3}\right)\delta_{ij}\delta_{kl}+(\bar{G}+\bar{G}_{c})\delta_{ik}\delta_{jl}+(\bar{G}-\bar{G}_{c})\delta_{il}\delta_{jk},\label{eq:nonlinear_strain_stiffness}\\
	D_{ijkl}^{e}&=\left(\bar{L}-\frac{2\bar{H}}{3}\right)\delta_{ij}\delta_{kl}+(\bar{H}+\bar{H}_{c})\delta_{ik}\delta_{jl}+(\bar{H}-\bar{H}_{c})\delta_{il}\delta_{jk}.\label{eq:nonlinear_curvature_stiffness}
\end{align}
$\bar{K}$, $\bar{G}$, $\bar{G}_{c}$, $\bar{L}$, $\bar{H}$ and $\bar{H}_{c}$ are the nonlinear analogues of the corresponding terms in linear Cosserat elasticity. If we set $n=0$, we recover classical linear Cosserat elasticity without any consideration of density dependence. We can use the same calibration relationships developed in \citet{Collins-Craft2020} to obtain the values of $\bar{G}_{c}$, $\bar{H}$ and $\bar{H}_{c}$, under the same assumption of $\bar{L}=0$. Non-zero values of $n$ allow us to recover the observed dependence of stiffness on the density of the material \citep{Hardin1972}. From this potential we are able to derive the elastic stress and couple stress, the chemical potential and thermodynamic pressure, as well as the breakage energy by following the standard procedures detailed in Appendix~\ref{sec:model_derivation}.\\\\
We suppose that the yield surface of the material is given by
\begin{equation}
	y=\left(\sqrt{\frac{E_{B}}{E_{c}}}(1-B)-\zeta\chi\right)^{2}+\left(\frac{q}{Mp^{e}+\phi(1-B)c}\right)^{2}-1\leq0,\label{eq:y_mix}
\end{equation}
where $M$ is the stress ratio in the $p-q$ plane given by a function specified below, $E_{c}$ is the critical breakage energy, $c$ is the cohesion of the material, and $\zeta$ is a material parameter that controls the potential to dilate, and ranges from $0$ (no dilation) to $1$ (maximally dilating). We note that in the case where we set $c=0$ and $\zeta=0$, we recover the yield surface given in \citet{Collins-Craft2020}.\\\\
We postulate that the plastic evolution rules of the system are given by
\begin{align}
	\dot{B}&=\lambda\left\langle F\right\rangle\frac{2(1-B)}{\sqrt{E_{B}E_{c}}}\cos^{2}(\omega),\label{eq:B_dot}\\
	\dot{\phi}^{p}&=\lambda F\frac{2\phi(1-B)}{p^{e}}\sqrt{\frac{E_{B}}{E_{c}}}\sin^{2}(\omega),\label{eq:phi_p_dot}\\
	\dot{\gamma}_{s}^{p}&=\lambda\frac{2q}{(Mp^{e}+\phi(1-B)c)^{2}},\label{eq:gamma_s_p_dot}
\end{align}
where $\langle\cdot\rangle=(\cdot+|\cdot|)/2$ are Macaulay brackets, and $F$ is a function given by
\begin{equation}
	F=\sqrt{\frac{E_{B}}{E_{c}}}(1-B)-\zeta\chi,\label{eq:F}
\end{equation}
and $\omega$ is a coupling angle that allocates dissipation between plastic volumetric straining and grain breakage given by
\begin{equation}
	\omega=\frac{\pi}{2}(1-\chi\mathcal{H}(F)),\label{eq:omega}
\end{equation}
where $\mathcal{H}(\cdot)$ is the Heaviside step function, defined such that
\begin{equation}
	\mathcal{H}(x)=\begin{cases}
		0 & \textrm{if }x<0,\\
		1 & \textrm{if }x\geq0.
	\end{cases}\label{eq:Heaviside_definition}
\end{equation}
We take these evolution rules from \citet{Tengattini2016}, who demonstrated the model's ability to represent crushable granular soils that can experience dilation. We generalise their rules slightly to include cohesion in the expression for the shear strain rate invariant, allowing us to model granular rocks in addition to granular soils. We can use \eqref{eq:phi_dot_p} of Appendix~\ref{sec:thermodynamic_admissibility} to transform \eqref{eq:phi_p_dot} into an expression for $\dot{\varepsilon}_{v}^{p}$, as well as the derivative of \eqref{eq:gamma_s_p_dot} with respect to the plastic deviatoric strain and curvature rates to deduce the flow rule for the complete set of plastic strain and curvature rates:
\begin{align}
	\dot{\gamma}_{ij}^{p}&=\lambda\left[F\frac{2(1-B)}{3p^{e}}\sqrt{\frac{E_{B}}{E_{c}}}\sin^{2}(\omega)\delta_{ij}+\frac{2}{(Mp^{e}+\phi(1-B)c)^{2}}\left(h_{1}^{\star}s_{ij}^{e}+h_{2}^{\star}s_{ji}^{e}\right)\right],\label{eq:gamma_p_dot}\\
	\dot{\kappa}_{ij}^{p}&=\lambda\frac{2}{[\ell(Mp^{e}+\phi(1-B)c)]^{2}}\left(h_{3}^{\star}m_{ij}+h_{4}^{\star}m_{ji}\right).\label{eq:kappa_p_dot}
\end{align}
As is well-known, densely-packed sands dilate under shear, and the function $F$ allows the model to express how the behaviour changes as the density and GSD evolves. When $F<0$ the system dilates due to grain rearrangement without grain breakage \textit{i.e.} the rate of breakage $\dot{B}=0$ and the plastic rate of the solid fraction $\dot{\phi}^{p}<0$. We refer to this as the dilative side of the critical state line. When $F>0$ grains break and the solid fraction increases \textit{i.e.} $\dot{B}>0$ and $\dot{\phi}>0$. We will refer to this as the compactive side of the critical state line. When $F=0$ we are in a state of zero breakage growth and isochoric plastic straining, \textit{i.e.} the critical state. In this model, the critical state appears at a lower confining stress than the peak deviatoric stress. The mathematical form of \eqref{eq:omega} allows us to guarantee all the volumetric straining goes to solid fraction changes when $F<0$, while smoothly varying the allocation between grain breakage and pore collapse when $F>0$. This is a point of contrast with the model specified in \citet{Collins-Craft2020}, where $\omega$ is a fixed material constant that does not evolve in time, and dilation cannot occur.\\\\
We take $M$ to be dependent on the position of the state relative to the critical state by:
\begin{equation}
	M=M_{0}[1+\zeta\mathcal{H}(-F)F].\label{eq:M_variable}
\end{equation}
Thus, where our system is dense $M$ is larger, while looser systems have a correspondingly lower value of $M$, matching experimental observations \citep{Bolton1986,Wichtmann2016}. $M_{0}$ is given by another function
\begin{equation}
	M_{0}=\frac{3\sin(\varphi)}{\sqrt{3}\cos(\beta)-\sin(\beta)\sin(\varphi)},\label{eq:M_0_expression}
\end{equation}
where $\varphi$ is the internal friction angle (a material parameter with $\varphi\geq0$) and $\beta$ is the Lode angle. Experimental and theoretical results have shown that the Lode angle plays a role in the localisation behaviour of rocks such as sandstones \citep{Couture2022,Couture2023}. In the Cosserat continuum, this angle may be defined as:
\begin{equation}
	\beta=\frac{1}{3}\arcsin\left(\frac{27}{2}\frac{\det(s_{ij})}{{q}^{3}}\right),\label{eq:Lode_angle}
\end{equation}
and classically lies between $-\pi/6$ and $\pi/6$ \citep{Papamichos2010}. We use the definition such that in triaxial compression the Lode angle is equal to $\pi/6$.\\\\
Finally, we must make a constitutive assumption about the evolution of the solid density. Most commonly, solids are assumed to have a fixed density, but here we consider its inevitable change driven by elastic volumetric straining. We follow \citet{Alaei2021} in specifying
\begin{equation}
	\dot{\rho}_{s}=\rho_{s}\chi\dot{\varepsilon}_{v}^{e}.\label{eq:dot_rho_s}
\end{equation}
The model is now fully specified, with the detailed derivations and the proof of the non-negativity of the rate of dissipation (as required by the second law of thermodynamics) presented in Appendix~\ref{sec:model_derivation}. We note here in passing that by setting $n=0$, $c=0$, $\zeta=0$ and $\chi=1$ we can recover the model specified in \citet{Collins-Craft2020} for $\bar{\omega}=0^{\circ}$ and loading at constant volume. For any other system values or different load conditions, we will recover different behaviour due to the changes that we have made to the volumetric parts of the model.
\section{Parametric study}\label{sec:parametric_study}
\subsection{Calibration}\label{sec:calibration}
\begin{figure}[H]
	\centering
	\includegraphics{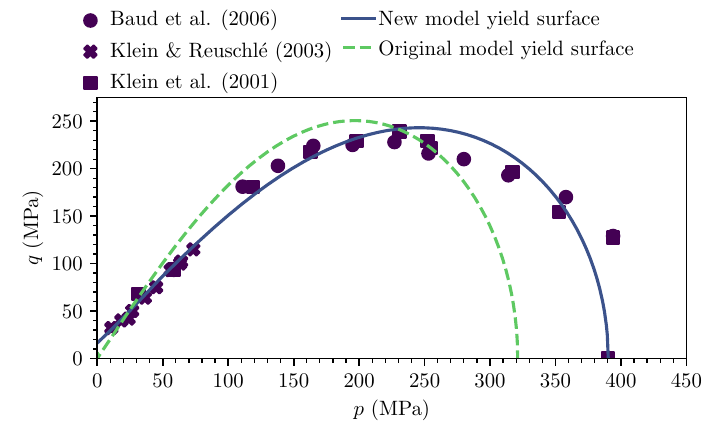}
	\caption{Yield surfaces of the model presented in this paper and the model presented in \citet{Collins-Craft2020}, and the experimental values from \citet{Klein2001,Klein2003,Baud2006} used to calibrate the relevant parameters.}
	\label{fig:calibrated_yield_surface}
\end{figure}

\begin{figure}[H]
	\centering
	\includegraphics{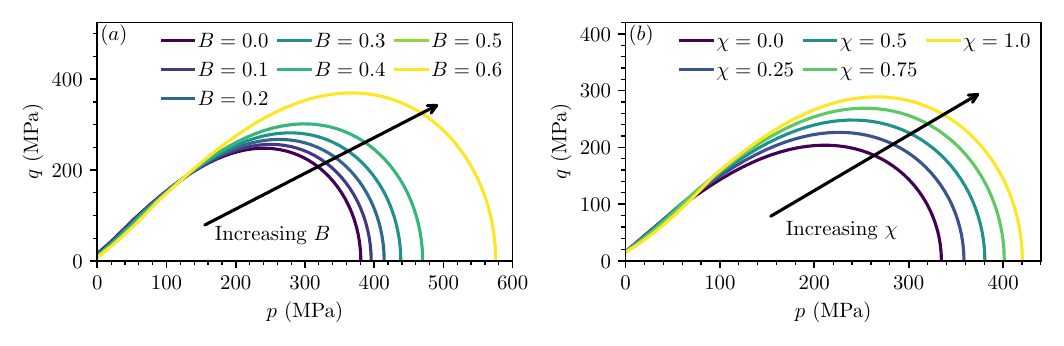}
	\caption{The changes in the yield surface in $p-q$ space as (a) the breakage is varied from $0$ to $0.6$ in increments of $0.1$ and as (b) the relative solid fraction is varied from $0$ to $1$ in increments of $0.25$.}
	\label{fig:calibrated_yield_surface_varying_B_and_chi}
\end{figure}

\begin{figure}[H]
	\centering
	\includegraphics{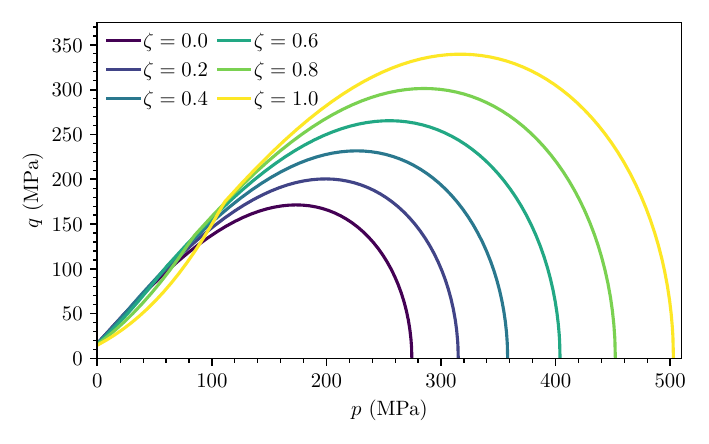}
	\caption{The changes in the yield surface in $p-q$ space as $\zeta$ is varied from $0$ to $1$ in increments of $0.2$.}
	\label{fig:zeta_sensitivity_yield_surface}
\end{figure}

We calibrate the proposed model against Bentheim sandstone, which has been the subject of extensive experimental studies, and has a large-grained composition of approximately 95\% quartz \citep{Klein2001}, which is ideal for demonstrating the capabilities of the model. We assume samples to start in an unbroken state ($B=0$) and choose an initial relative solid fraction of $\chi_{0}=0.5$, so that in combination with the upper and lower bound given below in Table~\ref{tab:parameter_values}, the initial solid fraction $\phi_{0}=0.775$. We select the potential to dilate $\zeta=0.5$ as the middle value of its possible range. We then obtain the other parameter values from the literature, either directly or via back-calibration using our chosen initial values for the state variables and the parameter $\zeta$.\\\\
The minimum grain $d_{\textrm{min}}=1$~$\mu$m size is chosen following arguments in \citet{Kendall1978} and \citet{Buscarnera2012}. The maximum grain size $d_{\textrm{Max}}$ is as reported for Bentheim sandstone in \citet{Klein2001}. $\alpha$ is chosen in line with studies of the GSDs of mature faults \citep{Sammis1986,Sammis1987,Sammis1989,An1994}. The derived granulometric parameters are then calculated via \eqref{eq:theta_gamma_def}, \eqref{eq:theta_kappa_def} and \eqref{eq:theta_I_def}. The lower and upper limit leading terms are estimates of the plausible range based on the reported porosities in \citet{Klein2001} (22.8\%) and \citet{Noel2021} (24.0\%). The upper and lower solid fraction bounds are taken directly from \citet{Cil2020}, in the absence of any experimental evidence on what the appropriate values would be for this sandstone. The density nonlinearity index is chosen to be 3, in line with \citet{Alaei2022}. In general this index can be calibrated against standard triaxial compression tests. The nonlinear bulk and shear moduli are taken from the values reported in \citet{Noel2021}, divided by their reported solid fraction raised to the density nonlinearity index. These can be obtained either by fitting to the load-displacement curve of standard triaxial compression tests, or using p- and s-wave velocities. We use the three-dimensional kinematic Cosserat model, which determines the values of $g_{i}^{\star}$ and $h_{i}^{\star}$. $\bar{L}$ is set to zero as we do not expect the torsional effects to play any role in our intended application. We then use the calibration relationships derived in \citet{Collins-Craft2020}, which give the values of $\bar{G}_{c}$, $\bar{H}$ and $\bar{H}_{c}$ by
\begin{align}
	\bar{G}_{c}&=\frac{3\bar{G}}{2(h_{1}^{\star}-h_{2}^{\star})},\label{eq:Gc_calibration}\\
	\bar{H}&=\frac{3\bar{G}}{2(h_{3}^{\star}+h_{4}^{\star})},\label{eq:H_calibration}\\
	\bar{H}_{c}&=\frac{3\bar{G}}{2(h_{3}^{\star}-h_{4}^{\star})},\label{eq:Hc_calibration}
\end{align}
The density of the solid material is taken by dividing the reported bulk density in \citet{Noel2021} by their reported solid fraction, two quantities that can be measured by weighing samples when both dry and wet. The solid fraction can also be obtained by x-ray $\mu$-CT scans where available.\\\\
This leaves us three remaining constants to calibrate using experiments. $E_{c}$ is generally obtained by performing an isotropic volumetric compaction and observing the point at which the grains start to break, then re-arranging \eqref{eq:y_mix} suitably. In our case we take the value given in \citet{Klein2001} (as well as our previously established material parameters) and calculate the elastic strains (and associated density and solid fraction) for their reported crushing pressure of 390~MPa, noting that for their samples of Bentheim sandstone they report an initial solid fraction of 0.772 (corresponding to $\chi_{0}=0.44$), so we use this value for the calibration, rather than our assumed initial value of 0.775 (corresponding to $\chi_{0}=0.5$). Then, having obtained the set of state variables at the crushing pressure (and hence $\chi$ and $E_{B}$), we obtain $E_{c}$ by
\begin{equation}
	E_{c}=\frac{E_{B}(1-B)^{2}}{(1+\zeta\chi)^{2}}.\label{eq:Ec_from_crushing_pressure}
\end{equation}
$\varphi$ can be obtained by observing the ratio between $p$ and $q$ in a triaxial test when the system enters the critical state, while $c$ can be determined by using lines-of-best fit from triaxial compression tests done at very low confining stresses. Here, we obtain $\varphi$ and $c$ by a standard least squares fitting procedure that minimises the norm of $y$ evaluated at each experimental data point. We use the results of the triaxial compression tests reported for Bentheim sandstone in \citet{Klein2001,Klein2003,Baud2006} as the source of the calibration data, taking the values reported at ``the onset of dilatancy'', and excluding the isotropic compaction tests. The final set of obtained calibrated values is contained in Table~\ref{tab:parameter_values}.
\begin{table}[H]
	\centering
	\begin{tabular}{rlll}
		\hline
		\multicolumn{4}{c}{Basic granulometric parameters}\\ \hline				
		Minimum grain size & $d_{\textrm{min}}$ & $0.001$ & mm\\
		Maximum grain size & $d_{\textrm{Max}}$ & $0.5$ & mm\\
		Ultimate GSD exponent & $\alpha$ & 2.6 & \\ \hline
		\multicolumn{4}{c}{Derived granulometric parameters}\\ \hline
		First grading index & $\theta_{\gamma}$ & $0.818$ & \\
		Second grading index & $\theta_{\kappa}$ & $0.901$ & \\
		Third grading index & $\theta_{I}$ & $0.919$ & \\ \hline
		\multicolumn{4}{c}{Porosimetric parameters}\\ \hline
		Upper limit leading term & $\alpha_{\textrm{upper}}$ & $0.2$ & \\
		Lower limit leading term & $\alpha_{\textrm{lower}}$ & $0.25$ & \\
		Upper solid fraction bound & $u_{\phi}$ & $0.12$ & \\
		Lower solid fraction bound & $l_{\phi}$ & $0.16$ & \\ \hline
		\multicolumn{4}{c}{Mechanical parameters}\\ \hline
		Density nonlinearity index & $n$ & $3$ & \\
		Nonlinear bulk stiffness & $\bar{K}$ & $19363$ & MPa \\
		Nonlinear shear stiffness & $\bar{G}$ & $15718$ & MPa \\
		Cosserat kinematic model $h_{i}$ parameters & $\{h_{1}^{\star},h_{2}^{\star},h_{3}^{\star},h_{4}^{\star}\}$ & $\{6/5,3/10,6/5,3/10\}$ & \\
		Cosserat kinematic model $g_{i}$ parameters & $\{g_{1}^{\star},g_{2}^{\star},g_{3}^{\star},g_{4}^{\star}\}$ & $\{8/9,-2/9,8/9,-2/9\}$ & \\
		First nonlinear Cosserat shear stiffness & $\bar{G}_{c}$ & $26197$ & MPa \\
		Nonlinear Cosserat torsion stiffness & $\bar{L}$ & $0$ & MPa \\
		Second nonlinear Cosserat shear stiffness & $\bar{H}$ & $15718$ & MPa \\
		Third nonlinear Cosserat shear stiffness & $\bar{H}_{c}$ & $26197$ & MPa \\
		Unstressed solid mass density & $\rho_{s}^{\star}$ & $2.645\times10^{-3}$ & g/mm$^{3}$ \\
		Critical breakage energy & $E_{c}$ & $3.32$ & MPa \\
		Internal friction angle & $\varphi$ & $0.583$ & \\
		Cohesion & $c$ & $21.5$ & MPa \\
		Potential to dilate & $\zeta$ & $0.5$ & \\ \hline
	\end{tabular}
	\caption{Calibrated parameter values for the model developed in this paper.}
	\label{tab:parameter_values}
\end{table}
To demonstrate the capabilities offered by the proposed model, we compare it with the model detailed in \citet{Collins-Craft2020}, which entails calibrating that model against the same data. We note that the model in \citet{Collins-Craft2020} does not possess certain parameters or their equivalents (namely, the porosimetric terms, the density nonlinearity index, the unstressed solid mass density and the cohesion). In the newly proposed model, $M$ and $\omega$ are functions of the state, the parameter $\varphi$ is calibrated against data and the parameter $\zeta$ is chosen to be in the middle of its possible range. In the previous model from \citet{Collins-Craft2020}, $\bar{M}$ is a parameter that we calibrate against data and $\bar{\omega}$ is a parameter chosen to be in the middle of its possible range. We add a bar to distinguish between the usages of these quantities as variables and parameters, and further distinguish $\bar{\omega}$ by reporting its value in degrees, while the variable $\omega$ is in radians.\\\\
We once again obtain the relevant parameters ($\bar{M}$ and $E_{c}$) by minimising the norm of $y$ against the experimental data. However, due to the reduced capacity of the previous model to fully fit the experimental data, we fit over a reduced range of points so that in the region where we conduct our simulations, the quality of the fit is reasonable. We use only the experimental data points with $p$ less than 350~MPa to fit the yield surface, and as a consequence at higher stresses there is a severe mismatch between the model yield surface and the experimental values. Further, we set the bulk stiffness $K$ and shear stiffness $G$ to have the same values as the effective nonlinear stiffness at $200$~MPa. The full set of calibrated values for the model presented in \citet{Collins-Craft2020} are given in Table~\ref{tab:original_parameter_values}.
\begin{table}[H]
	\centering
	\begin{tabular}{rlll}
		\hline
		\multicolumn{4}{c}{Basic granulometric parameters}\\ \hline				
		Minimum grain size & $d_{\textrm{min}}$ & $0.001$ & mm\\
		Maximum grain size & $d_{\textrm{Max}}$ & $0.5$ & mm\\
		Ultimate GSD exponent & $\alpha$ & $2.6$ & \\ \hline
		\multicolumn{4}{c}{Derived granulometric parameters}\\ \hline
		First grading index & $\theta_{\gamma}$ & $0.818$ & \\
		Second grading index & $\theta_{\kappa}$ & $0.901$ & \\
		Third grading index & $\theta_{I}$ & $0.919$ & \\ \hline
		\multicolumn{4}{c}{Mechanical parameters}\\ \hline
		Bulk stiffness & $K$ & $19.9\times10^{3}$ & MPa \\
		Shear stiffness & $G$ & $16.2\times10^{3}$ & MPa \\
		Cosserat kinematic model $h_{i}$ parameters & $\{h_{1}^{\star},h_{2}^{\star},h_{3}^{\star},h_{4}^{\star}\}$ & $\{6/5,3/10,6/5,3/10\}$ & \\
		Cosserat kinematic model $g_{i}$ parameters & $\{g_{1}^{\star},g_{2}^{\star},g_{3}^{\star},g_{4}^{\star}\}$ & $\{8/9,-2/9,8/9,-2/9\}$ & \\
		First Cosserat shear stiffness & $G_{c}$ & $27.0\times10^{3}$ & MPa \\
		Cosserat torsion stiffness & $L$ & 0 & MPa \\
		Second Cosserat shear stiffness & $H$ & $16.2\times10^{3}$ & MPa \\
		Third Cosserat shear stiffness & $H_{c}$ & $27.0\times10^{3}$ & MPa \\
		Critical breakage energy & $E_{c}$ & $2.12$ & MPa \\
		Slope of the critical state line in the $p-q$ plane & $\bar{M}$ & $2.08$ & \\
		Coupling angle & $\bar{\omega}$ & $45$ & $^{\circ}$\\ \hline
	\end{tabular}
	\caption{Calibrated parameter values for the model presented in \citet{Collins-Craft2020}.}
	\label{tab:original_parameter_values}
\end{table}
The yield surfaces obtained using the calibrated values (with $\phi=0.772$ for the model presented in this paper) against the experimentally obtained values are shown in \figref{fig:calibrated_yield_surface}.
\figref{fig:calibrated_yield_surface} shows that the new model is able to fit the experimental data substantially better than the original model. The tests done at low confinement stresses clearly point towards Bentheim sandstone having cohesion, which the new model accommodates, while the original model cannot. Also visible in the experimental data is a slight non-convexity of the yield surface at low confining stresses, which only the new model can match. In general, we are not aware of any previously developed model that has been fitted to these data points that could recover this observation, and very often classical plasticity models insist on the convexity of the yield surface. This restriction can be lifted without affecting the non-negativeness of the rate of dissipation, as demonstrated in Appendix~\ref{sec:thermodynamic_admissibility}. As such, we conclude that the new model more accurately describes the initiation of plasticity than the original model, at low, intermediate and high ranges of confining stress.\\\\
We observe how the yield surface of the new model changes as we vary the breakage index and the relative solid fraction, in \figref{fig:calibrated_yield_surface_varying_B_and_chi}.
As the breakage value increases, the yield surface grows steadily larger and moves rightward, with decreasing shear strength before yield at lower confining stresses and increasing shear capacity at higher confining stresses. This is the same pattern as the for the model in \citet{Collins-Craft2020}, other than an absence of decrease of shear strength at low confining stresses. Similarly, as the solid fraction is increased the yield surface also enlarges, although the effect on the shear behaviour is much smaller than that of the breakage index.\\\\
We can also examine the effect of changing the parameter $\zeta$, the tendency to dilate, on the system, varying from $\zeta=0$ (minimal dilation) to $\zeta=1$ (maximal dilation), and shown in \figref{fig:zeta_sensitivity_yield_surface}.
The degree of non-convexity in \figref{fig:zeta_sensitivity_yield_surface} changes with $\zeta$. This is a necessary consequence of the system softening with dilation, in a manner analogous to the softening with increasing wetting in unsaturated soil mechanics, where non-convex yield surfaces are common \citep{Sheng2008}. The point at which the convexity of the yield surface changes does not require any special numerical treatment to deal with the nonsmooth behaviour, as both $\dot{B}$ and $\dot{\phi}^{p}$ will be zero at this point and so only the shear plastic flow component will be non-zero.
\subsection{Simulations and stability analysis of shear tests}\label{sec:element_simulations}
In order to understand the mechanical behaviour as well as the localisation tendency of the system we consider two loading cases, shearing at constant volume, which approximates the fast undrained shear that faults typically experience, and shearing at constant confining stress, which approximates the behaviour of the system under slow drained loading. Per \citet{Sulem2011} 200~MPa is a representative pressure for the depth at which seismic faulting typically occurs, so we choose this as our central value for the initial confining pressure in both load cases. We confine the system uniformly (that is $\gamma_{11}^{e}=\gamma_{22}^{e}=\gamma_{33}^{e}$ at the start of the simulation) and apply 0.2 strain in to the $\gamma_{12}$ and $\gamma_{21}$ entries, and simulate from this point while maintaining the appropriate boundary condition, using the numerical methods detailed in supporting~information. Once the results of each simulation have been obtained, a linear stability analysis is conducted where the system is perturbed and we study whether the perturbations grow or decay in time. We fix the perturbation to be horizontal \textit{i.e.} $\tilde{n}_{i}=\{1,0,0\}$, as analyses allowing the band orientation to vary have shown this to be the orientation with the fastest growth rate. An eigenvector analysis is also undertaken to qualitatively characterise the nature of the shear band. These procedures are detailed in supporting~information.
\subsubsection{Sensitivity to $\chi$}\label{sec:contstant_volume_chi_sensitivity}
\begin{figure}[H]
	\centering
	\includegraphics{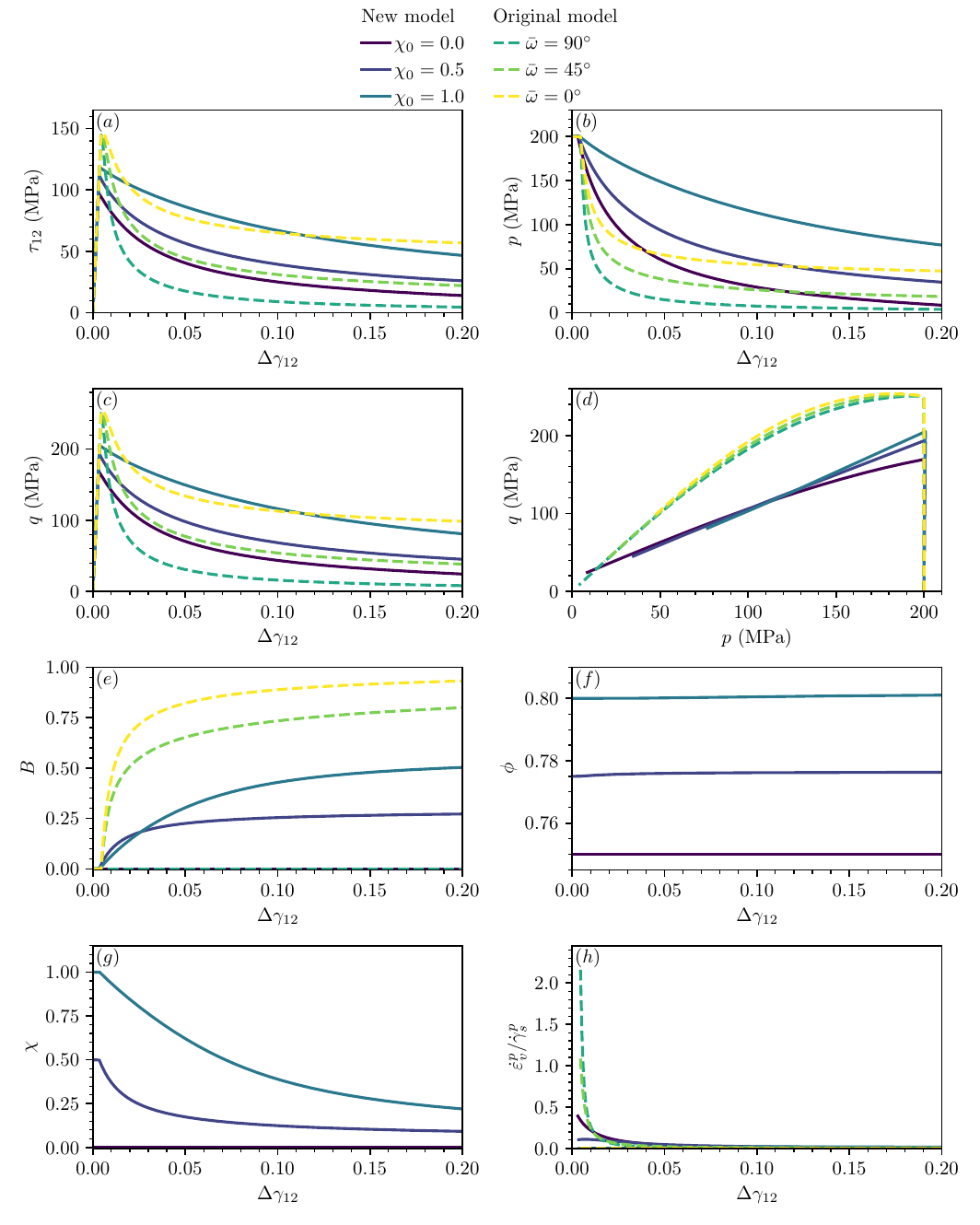}
	\caption{The results of simulations of shearing at constant volume with an initial confining stress of 200~MPa, using the model presented in this paper ($\chi\in\{0.0,0.5,1.0\}$) and the model presented in \citet{Collins-Craft2020} ($\bar{\omega}\in\{0^{\circ},45^{\circ},90^{\circ}\}$). (a) The shear stress $\tau_{12}$ against the increment of shear strain $\Delta\gamma_{12}$, (b) the mean stress $p$ against the increment of shear strain $\Delta\gamma_{12}$, (c) the deviatoric stress invariant $q$ against the increment of shear strain $\Delta\gamma_{12}$, (d) the deviatoric stress invariant $q$ against the mean stress $p$, (e) the breakage index $B$ against the increment of shear strain $\Delta\gamma_{12}$, (f) the solid fraction $\phi$ against the increment of shear strain $\Delta\gamma_{12}$, (g) the relative solid fraction $\chi$ against the increment of shear strain $\Delta\gamma_{12}$, and (h) the ratio of the plastic volumetric strain rate and the plastic shear strain rate invariant against the increment of shear strain $\Delta\gamma_{12}$. The quantities $\phi$ and $\chi$ do not exist in the original model, and hence only the new model is shown in subfigures (f) and (g).}
	\label{fig:chi_sensitivity_constant_volume_p_initial_200}
\end{figure}

\begin{figure}[H]
	\centering
	\includegraphics{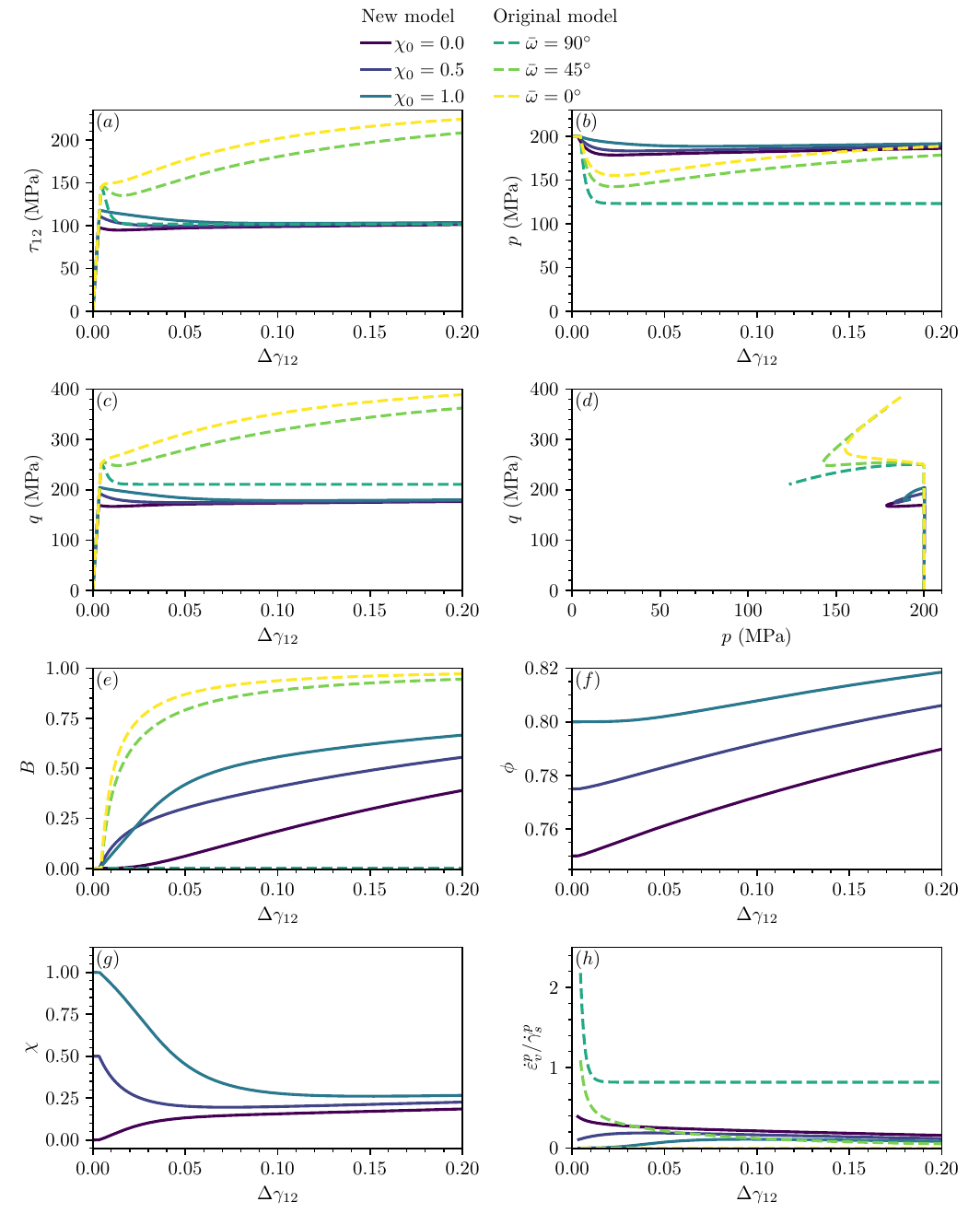}
	\caption{The results of simulations of shearing at constant confining stress of 200~MPa, using the model presented in this paper ($\chi\in\{0.0,0.5,1.0\}$) and the model presented in \citet{Collins-Craft2020} ($\bar{\omega}\in\{0^{\circ},45^{\circ},90^{\circ}\}$). (a) The shear stress $\tau_{12}$ against the increment of shear strain $\Delta\gamma_{12}$, (b) the mean stress $p$ against the increment of shear strain $\Delta\gamma_{12}$, (c) the deviatoric stress invariant $q$ against the increment of shear strain $\Delta\gamma_{12}$, (d) the deviatoric stress invariant $q$ against the mean stress $p$, (e) the breakage index $B$ against the increment of shear strain $\Delta\gamma_{12}$, (f) the solid fraction $\phi$ against the increment of shear strain $\Delta\gamma_{12}$, (g) the relative solid fraction $\chi$ against the increment of shear strain $\Delta\gamma_{12}$, and (h) the ratio of the plastic volumetric strain rate and the plastic shear strain rate invariant against the increment of shear strain $\Delta\gamma_{12}$. The quantities $\phi$ and $\chi$ do not exist in the original model, and hence only the new model is shown in subfigures (f) and (g).}
	\label{fig:chi_sensitivity_constant_confining_stress_p_initial_200}
\end{figure}

\begin{figure}[H]
	\centering
	\includegraphics{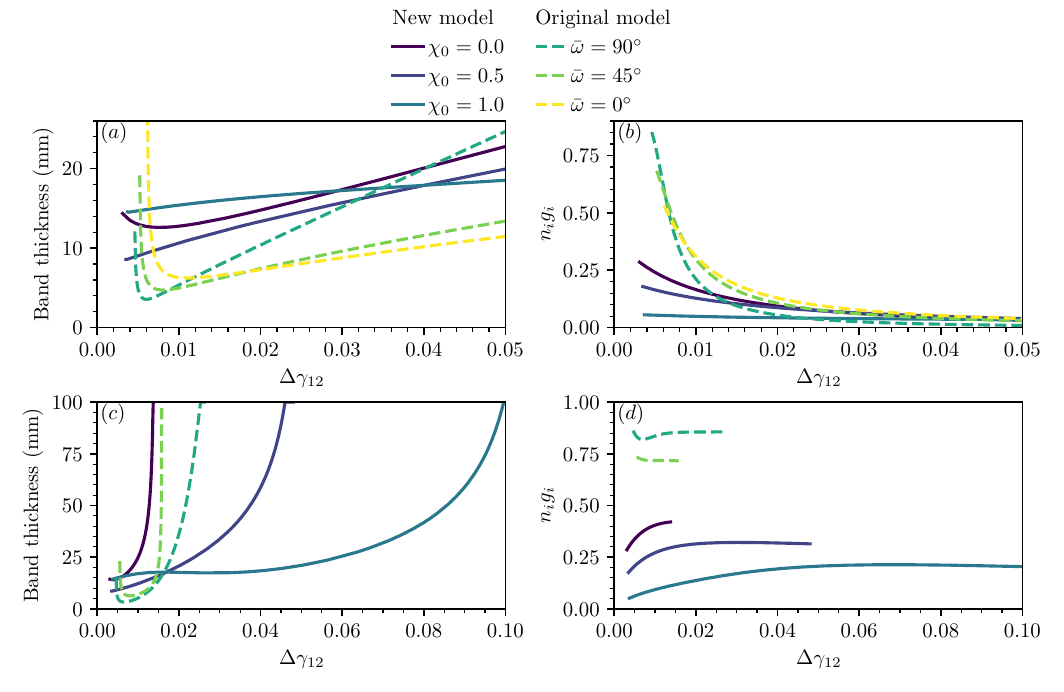}
	\caption{The results of the localisation analysis of simulations of shearing using the model presented in this paper ($\chi\in\{0.0,0.5,1.0\}$) and the model presented in \citet{Collins-Craft2020} ($\bar{\omega}\in\{0^{\circ},45^{\circ},90^{\circ}\}$). The constant confining stress simulations are truncated at $\Delta\gamma_{12}=0.1$ to better illustrate the delocalisation tendency. (a) The predicted band thickness against the increment of shear strain $\Delta\gamma_{12}$ for shear at constant volume, (b) the product of the orientation of the shear band with the eigenvector associated with the zero eigenvalue of the acoustic tensor against the increment of shear strain $\Delta\gamma_{12}$ for shear at constant volume, (c) the predicted band thickness against the increment of shear strain $\Delta\gamma_{12}$ for shear at constant confining stress, and (d) the product of the orientation of the shear band with the eigenvector associated with the zero eigenvalue of the acoustic tensor against the increment of shear strain $\Delta\gamma_{12}$ for shear at constant confining stress.}
	\label{fig:chi_sensitivity_localisation}
\end{figure}

Changing the initial value of the relative solid fraction $\chi_{0}$ allows us to study the effect of changing the initial volumetric behaviour of the new model. In the original model, we can change the values of $\bar{\omega}$ to similarly vary the volumetric behaviour. The results of these simulations are shown in \figref{fig:chi_sensitivity_constant_volume_p_initial_200} for constant volume shearing.
\figref{fig:chi_sensitivity_constant_volume_p_initial_200} demonstrates that under varying $\chi$ or $\bar{\omega}$, the trends remain broadly similar. As $\chi$ increases, the stress that can be sustained prior to yielding also increases slightly. By contrast, variations in $\bar{\omega}$ have no impact on the point at which yielding first occurs. For all values of $\chi_{0}$, the apparent softening experienced by the system is substantially less than for the corresponding system in the original model, although the general trends are similar ($\bar{\omega}$ favouring dissipation by plastic volumetric straining ($\bar{\omega}=90^{\circ}$) shows more apparent softening that $\bar{\omega}$ favouring dissipation by grain breakage ($\bar{\omega}=0^{\circ}$), likewise $\chi_{0}$ that favours volumetric compaction shows more apparent softening than $\chi_{0}$ favouring grain breakage). The original model displays substantially more grain breakage after a modest amount of shearing than the new model, as well as a greater initial tendency to plastic volumetric compaction relative to the plastic shear straining. The interaction between grain breakage and solid fraction in the new model allows the system to approach the critical state where the breakage state variable will stop evolving. The original model has a critical state at $B=1$, while in the new model the $F$ function allows the critical state to be approached at any value of $B$. Hence, the new model is able to much more accurately model the real physical behaviour of granular media, which can enter the critical state without requiring that the GSD has reached its ultimate state. This has been demonstrated in \citet{Tengattini2016}, where an extensive comparison with experimental results was undertaken. For the solid fraction and relative solid fraction variables that are only available within the new model, we see very little changes in the value of the solid fraction, as is expected for constant volume loading conditions. We see that the relative solid fraction declines for both $\chi_{0}=1$ and $\chi_{0}=0.5$, with both tending towards a common steady state (corresponding to the critical state), while the value of $\chi$ does not evolve at all for the $\chi_{0}=0$ system. This combination of loading conditions and initial state prevents any breakage occurring, in turn guaranteeing no changes to $\chi$. We also calculate the ratio of the plastic strain rate invariant terms, a quantity that is often controlled by a so-called dilatancy parameter. For both models this ratio evolves, tending towards zero, as required for the critical state, but the original model features a much higher initial value that decreases rapidly. We conclude that while both models demonstrate variable volumetric behaviour, and both favour compaction, the new model changes in a comparatively more stable way without extremely rapid increases in the value of the breakage index. Finally, the new model arrives at the yield surface at smaller values of stress and strain, due to the better fitting of the yield surface to the experimental results.\\\\
The sensitivity to changes in $\chi_{0}$ and $\bar{\omega}$ can likewise be studied for constant confining stress, shown in \figref{fig:chi_sensitivity_constant_confining_stress_p_initial_200}.
\figref{fig:chi_sensitivity_constant_confining_stress_p_initial_200} shows that in the original model, the stress paths of $\tau_{12}$, $p$ and $q$ diverge as $\bar{\omega}$ varies, while in the new model, they converge to similar values. This occurs despite substantial differences in the solid fraction and breakage index. As for the constant volume system, this behaviour may be attributed to the different critical states of the two models, with the new model able to approach its critical state at any value of $B$, while the original model can only do so as $B\rightarrow1$. The system with $\bar{\omega}=90^{\circ}$ experiences no breakage and so finds a steady state where the underlying state variables no longer evolve (notwithstanding that plastic volumetric strains continue to accumulate, so the system is not in the critical state). In comparison to the constant volume loading condition, we note that both models produce more grain breakage, but the difference is more significant in the new model, which also experiences a substantial increase in the solid fraction of the system, notwithstanding a substantial decrease in the \textit{relative} solid fraction of the system. Once again the two models converge towards a similar value of the ratio of plastic volumetric strain rate to plastic deviatoric strain rate, with the exception of the $\bar{\omega}=90^{\circ}$ system.\\\\
We also examine the localisation behaviour of the system in \figref{fig:chi_sensitivity_localisation}, using the methodology that is detailed in supporting~information. To briefly summarise, we perturb the displacement and micro-rotation fields and observe whether those perturbations grow or decay in time. We find the fastest growing (or slowest decaying) wavelength of the perturbation and take that to be twice the width of the shear band. We also evaluate the eigenvalues of the acoustic tensor to determine whether the shear band is compacting or dilating. We truncate the analysis to a shear strain increment of 0.05 as the linear stability analysis is only strictly valid up to the moment of localisation. Beyond this point, it can give only an indication of possible behaviour.
\figref{fig:chi_sensitivity_localisation} shows that in both load conditions the original model requires more shear in order to induce a localisation, and that the initial localisation prediction is followed by a rapid decline, and then growth. The new model by contrast demonstrates monotonic growth for all simulations except $\chi_{0}=0.0$ under constant volume, which demonstrates a very small decline in the predicted width after the initial localisation. Notably, in the new model, we observe that the dependence of the initial localisation width on the initial solid fraction seems to be non-monotonic in both load conditions. To our knowledge, the variation of the shear band width with the solid fraction has not been systematically studied experimentally, but in \citet{Alshibli2000} the inclination angle of shear bands in biaxial compression tests of sands did not vary monotonically with sample density, particularly at higher confining stress. We take these results to be in alignment with the proposed model prediction that $\phi$ does not play a strong role in initial shear band thickness, nor is there any particular guarantee of monotonicity between $\phi$ and the band thickness. The original model displays a greater initial tendency towards compaction in the shear band, but under constant volume conditions all systems trend steadily towards states of pure shear. The systems that favour grain breakage show less tendency towards compaction, with the difference being especially marked for the new model, where the $\chi_{0}=1.0$ system shows only a very light compactive tendency initially. The key difference between the two load conditions is that while the constant volume system delocalises slowly, the constant confining stress condition delocalises very rapidly. The amount of shear required to delocalise varies with the relative solid fraction, with denser systems requiring greater deformation. The $\chi_{0}=1.0$ system demonstrates an apparent plateau in the shear band thickness for some amount of shear before resuming the widening process. The original model produces slightly thinner bands that narrow further and then delocalise more rapidly. The most breakage-favouring system ($\bar{\omega}=0^{\circ}$) does not localise at all under this loading condition. Both models demonstrate tendency towards plastic compaction under this loading, with the original model showing a substantially stronger effect, as in the constant volume shearing load cases. However, for the range of shear strain in which the shear bands are predicted to exist, there does not appear to be any convergence towards a constant value, unlike in the constant volume case.
\subsubsection{Sensitivity to $p$}\label{sec:constant_volume_p_sensitivity}
\begin{figure}[H]
	\centering
	\includegraphics{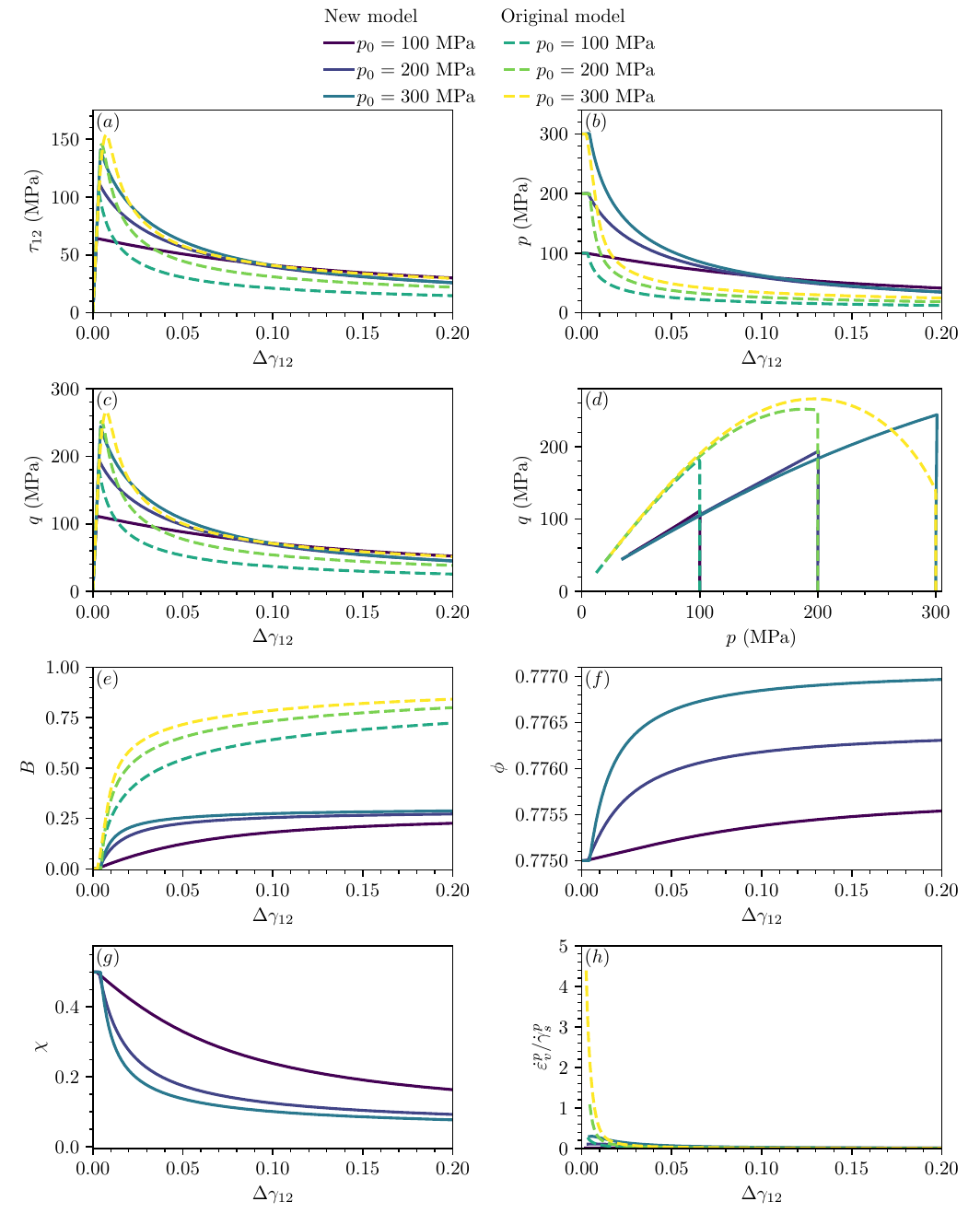}
	\caption{The results of simulations of shearing at constant volume, using the model presented in this paper and the model presented in \citet{Collins-Craft2020} (both with $p_{0}\in\{100,200,300\}$~MPa). (a) The shear stress $\tau_{12}$ against the increment of shear strain $\Delta\gamma_{12}$, (b) the mean stress $p$ against the increment of shear strain $\Delta\gamma_{12}$, (c) the deviatoric stress invariant $q$ against the increment of shear strain $\Delta\gamma_{12}$, (d) the deviatoric stress invariant $q$ against the mean stress $p$, (e) the breakage index $B$ against the increment of shear strain $\Delta\gamma_{12}$, (f) the solid fraction $\phi$ against the increment of shear strain $\Delta\gamma_{12}$, (g) the relative solid fraction $\chi$ against the increment of shear strain $\Delta\gamma_{12}$, and (h) the ratio of the plastic volumetric strain rate and the plastic shear strain rate invariant against the increment of shear strain $\Delta\gamma_{12}$. The quantities $\phi$ and $\chi$ do not exist in the original model, and hence only the new model is shown in subfigures (f) and (g).}
	\label{fig:p_sensitivity_constant_volume}
\end{figure}

\begin{figure}[H]
	\centering
	\includegraphics{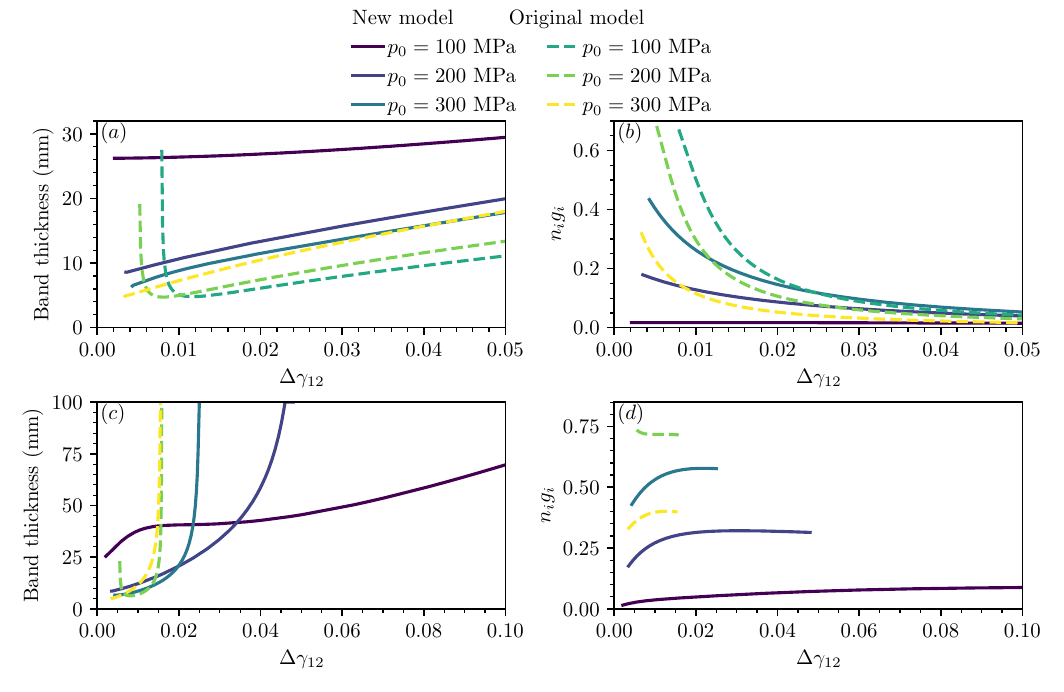}
	\caption{The results of the localisation analysis of simulations of shearing, using the model presented in this paper and the model presented in \citet{Collins-Craft2020} (both with $p_{0}\in\{100,200,300\}$~MPa). The constant confining stress simulations are truncated at $\Delta\gamma_{12}=0.1$ to better illustrate the delocalisation tendency. (a) The predicted band thickness against the increment of shear strain $\Delta\gamma_{12}$ for shear at constant volume, (b) the product of the orientation of the shear band with the eigenvector associated with the zero eigenvalue of the acoustic tensor against the increment of shear strain $\Delta\gamma_{12}$ for shear at constant volume, (c) the predicted band thickness against the increment of shear strain $\Delta\gamma_{12}$ for shear at constant confining stress, and (d) the product of the orientation of the shear band with the eigenvector associated with the zero eigenvalue of the acoustic tensor against the increment of shear strain $\Delta\gamma_{12}$ for shear at constant confining stress.}
	\label{fig:p_sensitivity_localisation}
\end{figure}

We may also consider the effect of varying $p_{0}$, the initial confining stress, on individual simulations. This represents changes in the burial depth of the fault. We show the load paths of the new model varying $p_{0}$ under constant volume shearing in \figref{fig:p_sensitivity_constant_volume}.
\figref{fig:p_sensitivity_constant_volume} shows that changes to the value of $p_{0}$ result in only modest changes to the quantitative behaviour of the system, with qualitative behaviour remaining very similar in this load case. In all cases, the systems demonstrate an apparent softening and an increase in $B$, with the apparent softening and value of $B$ both being greater with increases in initial confining pressure. The new model demonstrates much less breakage growth than the original model in this system, and the final value of $B$ shows relatively little sensitivity to $p_{0}$. The solid fraction increases slightly while the relative solid fraction declines slightly. The original model demonstrates much greater rates of breakage growth, accompanied by more dramatic apparent softening. However, the system with the highest initial confining stress ($p_{0}=300$~MPa) shows some post-yield hardening before softening. All systems show a tendency towards compaction, with this being greater in the old model than the new model, and at higher confining stresses than lower confining stresses. The tendency of all of the systems to head towards the critical state with no plastic volumetric straining is also clearly observable.
\begin{figure}[H]
	\centering
	\includegraphics{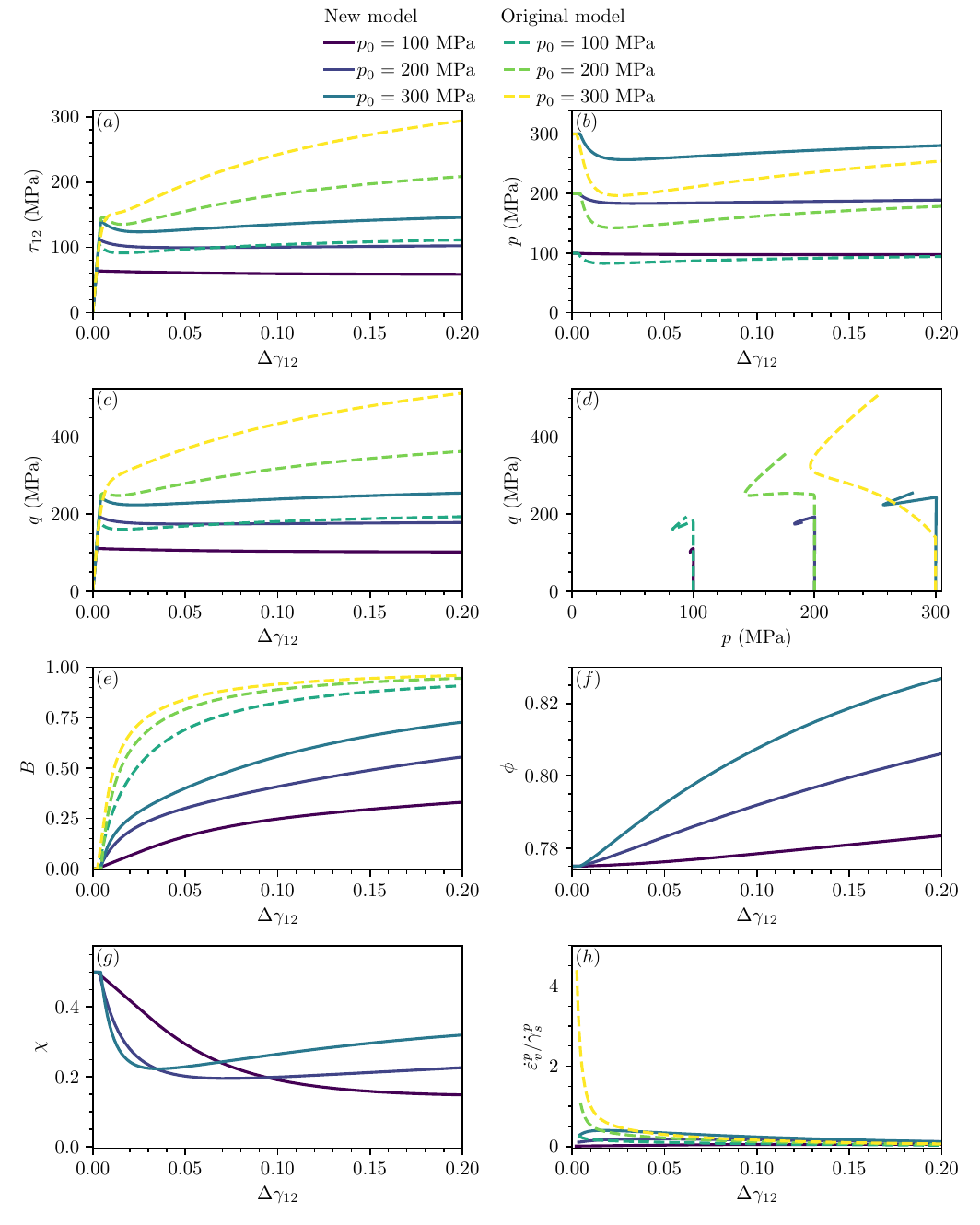}
	\caption{The results of simulations of shearing at constant confining stress, using the model presented in this paper and the model presented in \citet{Collins-Craft2020} (both with $p_{0}\in\{100,200,300\}$~MPa). (a) The shear stress $\tau_{12}$ against the increment of shear strain $\Delta\gamma_{12}$, (b) the mean stress $p$ against the increment of shear strain $\Delta\gamma_{12}$, (c) the deviatoric stress invariant $q$ against the increment of shear strain $\Delta\gamma_{12}$, (d) the deviatoric stress invariant $q$ against the mean stress $p$, (e) the breakage index $B$ against the increment of shear strain $\Delta\gamma_{12}$, (f) the solid fraction $\phi$ against the increment of shear strain $\Delta\gamma_{12}$, (g) the relative solid fraction $\chi$ against the increment of shear strain $\Delta\gamma_{12}$, and (h) the ratio of the plastic volumetric strain rate and the plastic shear strain rate invariant against the increment of shear strain $\Delta\gamma_{12}$. The quantities $\phi$ and $\chi$ do not exist in the original model, and hence only the new model is shown in subfigures (f) and (g).}
	\label{fig:p_sensitivity_constant_confining_stress}
\end{figure}
\figref{fig:p_sensitivity_constant_confining_stress} shows that under constant confining stress changing the value of $p_{0}$ can result in notable qualitative changes to the system, in contrast to the behaviour under constant volume shearing. For the original model, all systems demonstrate rapid growth in the value of the breakage index $B$, initial decreases in the value of $p$ followed by increases, and strong initial compactive tendencies that then move towards a state of pure shear. While the values of $q$ in the original model initially show some apparent softening for the lower confining stress simulations, they ultimately increase beyond their value at first yielding, while for the system under the greatest initial confining stress, this incrase is montonic. For the new model, we observe that the evolution of the system is slower than for the original model in all cases, with smaller (although still significant) increases in the value of $B$, as well as significant densification. While the system with the highest initial confining stress follows a qualitatively similar trajectory to the original model (albeit demonstrating some very slight decreases in the value of $q$ after yield before increasing again), the system at the lowest initial confining stress demonstrates almost no changes in the value of $p$ and $q$ after yielding. This indicates that the system is very close to the critical state, as can also be observed in the ratio of plastic volumetric compaction to plastic shearing, where the value remains very close to zero throughout the simulation. This qualitative difference in behaviour can only be achieved with the more realistic representation of the critical state in the new model.\\\\
We examine the changes in the system's localisation behaviour as $p_{0}$ is varied in \figref{fig:p_sensitivity_localisation}.
In \figref{fig:p_sensitivity_localisation} for both load cases the initial localisation width decreases monotonically with confining pressure, but the greater initial confining pressure induces faster growth. The constant confining stress systems all rapidly delocalise, with the exception of the lowest confining stress system ($p_{0}=100$~MPa), which does not localise under these load conditions in the original model, and grows steadily in width in the new model. Once again, all systems indicate that they support a compacting shear band, with the systems all evolving towards pure shear bands in the case of shearing under constant volume (noting that in the new model the least-confined system is very close to pure shear from the start of the localisation). In the constant confining stress case, the systems all indicate compacting shear bands, with the degree of compaction increasing monotonically with the confining stress. While almost all the systems indicate a small decrease in the ratio of compaction to shearing before completely delocalising, the system with $p_{0}=100$~MPa shows a gradual increase in the ratio of compaction to shearing in the shear band.
\subsection{Finite element simulations}\label{sec:finite_element_simulations}
\begin{figure}[H]
	\centering
	\includegraphics{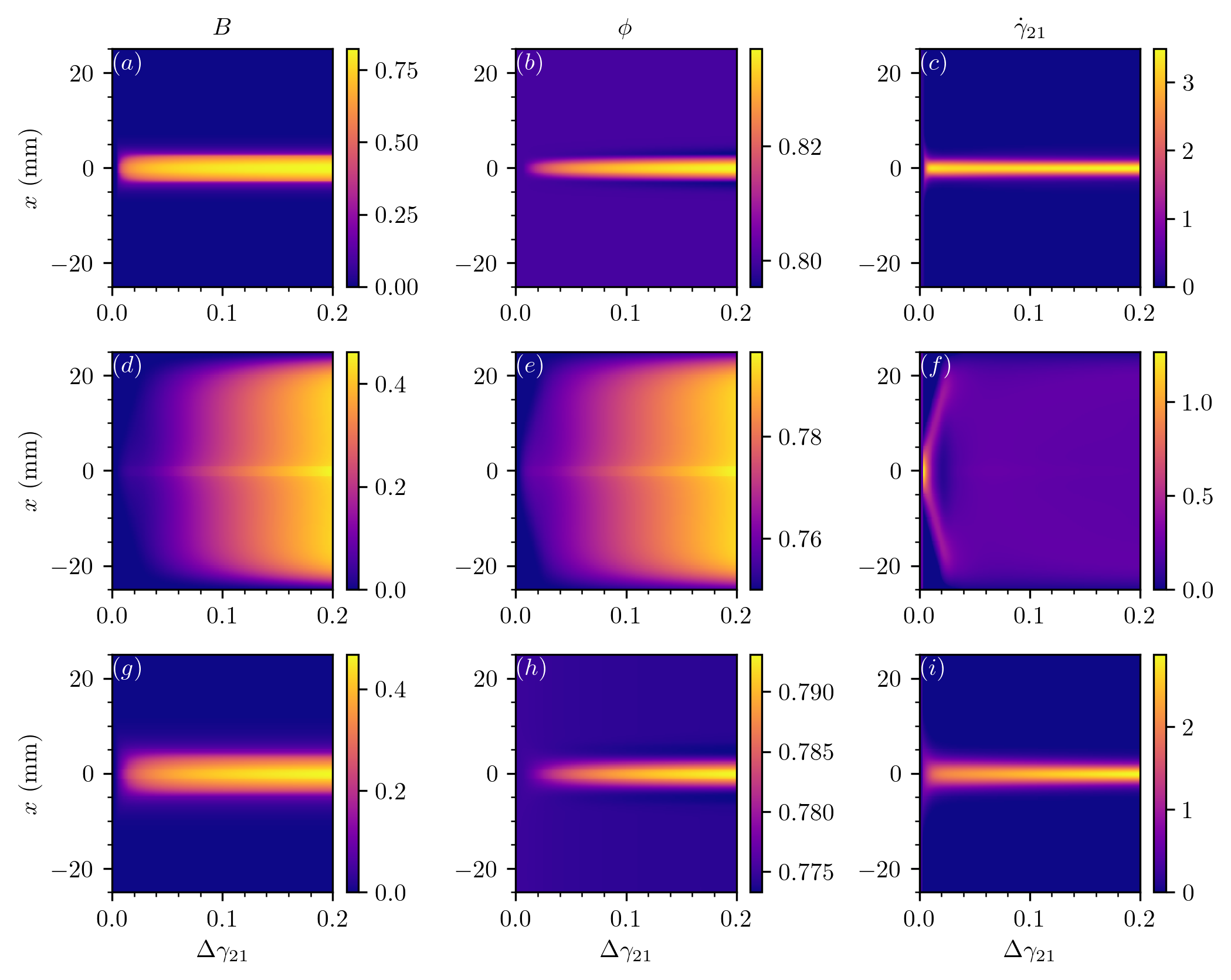}
	\caption{Spatiotemporal plots of three finite element simulations. The first column (subfigures (a), (d), (g)) are the values of $B$, the second column (subfigures (b), (e), (h)) are the values of $\phi$ while the third column (subfigures (c), (f), (i)) are the values of $\dot{\gamma}_{21}$. The first row (subfigures (a), (b), (c)) represent the constant volume simulation with $\chi_{0}=1.0$ and $p_{0}=200$~MPa, the second row (subfigures (d), (e), (f)) the constant confining stress simulation with $\chi_{0}=0.0$ and $p_{0}=200$~MPa and the third row (subfigures (g), (h), (i)) the constant volume simulation with $p_{0}=100$~MPa and $\chi_{0}=0.5$.}
	\label{fig:spatio_temporal_summary}
\end{figure}

Formally, the linear stability analysis is only valid up until the moment of localisation, as an assumption of the analysis is a bifurcation from a homogeneous to a non-homogeneous state of deformation. Thus, the trends observed after this point can at best be taken as indicative of the overall behaviour. In order to obtain a more exact analysis of the post-localisation behaviour, we use instead a method that does not require any assumption of homogeneity, namely the finite element method. We set up the finite element simulations as described in supporting information based on the Numerical Geolab framework \citep{Stefanou2024}, and consider the same variation in the loading conditions as well as $\chi_{0}$ and $p_{0}$ as for the linear stability analysis. All systems have zero horizontal and vertical displacements fixed on the bottom of the system and sufficient horizontal displacement on the top boundary to cause a homogeneous strain of 0.2, with the appropriate initial isotropic confining stress applied throughout. The constant volume simulations have zero vertical displacement fixed at the top while the constant confining stress simulations have a constant vertical stress applied.\\\\
We construct spatiotemporal plots of the evolution of the key variables to demonstrate the qualitative behaviour in \figref{fig:spatio_temporal_summary}.
In \figref{fig:spatio_temporal_summary}, we show the results of three simulations that represent the three classes of behaviour that can be observed over the full variation of $p_{0}$, $\chi_{0}$ and the loading conditions. The first class (shown in the first row of subfigures) is localisation into a relatively homogeneous band width where the rate of shear straining shows an increasing concentration at the centre of the band, suggesting a transition from continuous to discontinuous bifurcation of the system \citep{Rice1980}, where all deformation is accommodated in the band and elastic unloading outside the band becomes possible. The second class (shown in the second row of subfigures) is localisation into a shear band that then begins to delocalise. Here we can see that the shear strain rate decreases dramatically in the centre of the band and two zones featuring the most intense shearing move outwards from the centre until reaching the boundary. At the point at which the influence of the boundaries becomes significant, the system shears in a near-homogeneous fashion accompanied by steady GSD changes and densification throughout. The third class (shown in the third row of subfigures) is localisation into a band featuring increasing grain breakage and densification (similar to the first class), but accompanied by the formation of regions of dilation outside the band.
\section{Discussion}\label{sec:discussion}
\begin{figure}[H]
	\centering
	\includegraphics{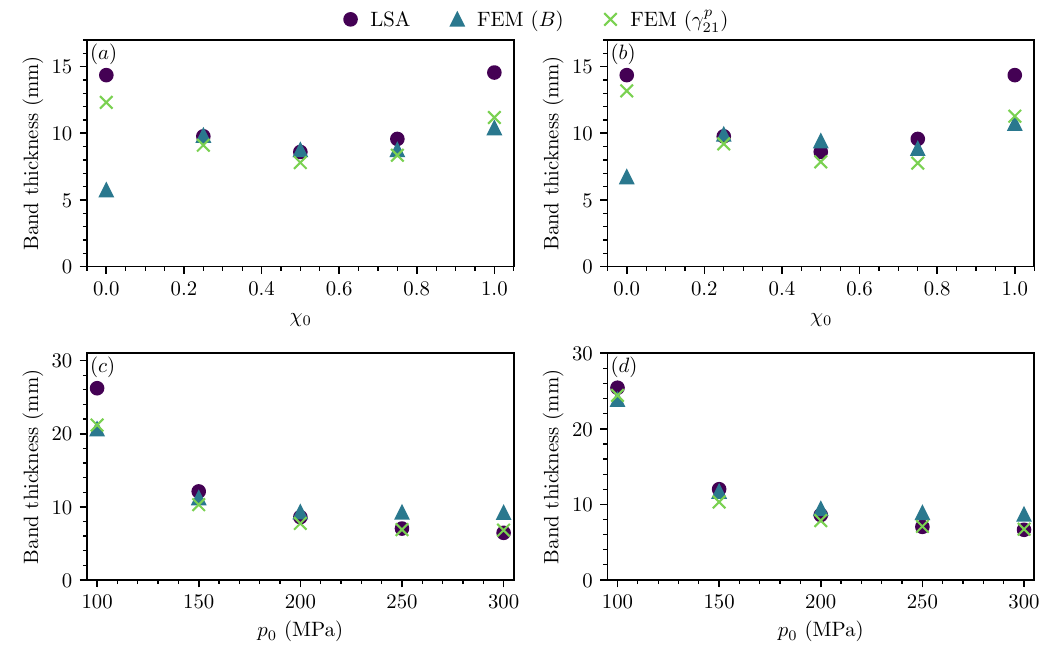}
	\caption{The predicted shear band widths from the LSA, and those inferred from the FEM simulations based on fitting on the spatial distribution of $B$ and $\gamma_{21}^{p}$. (a) The initial shear band widths when shearing under constant volume and varying $\chi_{0}$, (b) the initial shear band widths when shearing under constant confining stress and varying $\chi_{0}$, (c) the initial shear band widths when shearing under constant volume and varying $p_{0}$, and (d) the initial shear band widths when shearing under constant confining stress and varying $p_{0}$.}
	\label{fig:LSA_vs_FEM_comparison}
\end{figure}

\begin{figure}[H]
	\centering
	\includegraphics{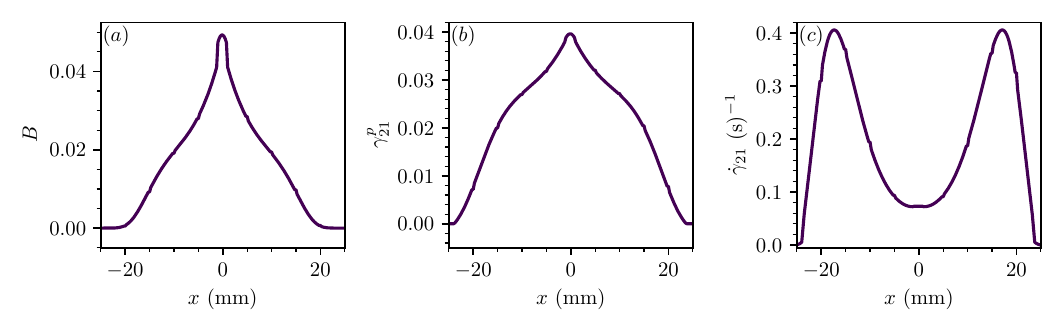}
	\caption{The spatial distributions of the some key variables at homogeneous shear strain $\Delta\gamma_{21}=0.025$ for the simulation with $\chi_{0}=0.0$ under constant confining stress (a) The spatial distribution of $B$, (b) the spatial distribution of $\gamma_{21}^{p}$, and (c) the spatial distribution of $\dot{\gamma}_{21}$.}
	\label{fig:illustrative_spatial_distribution_constant_confining_stress}
\end{figure}

We have shown in \secref{sec:element_simulations} and \secref{sec:finite_element_simulations} that changing the model parameters and state variables changes the physical characteristics of the system by modifying the way in which the compaction and grain crushing coevolve and potentially compete with dilation. This includes the shape of the resulting yield surface, where the model can accommodate cohesion $c$ that depends on the extent of the grain crushing $B$, classical convex yield surfaces that are able to capture the increasing amount of energy required to crush grains as compaction continues, as well as non-convex yield surfaces that arise in the proposed model as a consequence of dilation and reductions in the solid fraction $\phi$ at lower confining stresses. This last factor allows for a substantially better match to experimental data on Bentheim sandstones than was possible with the model that we developed in \citet{Collins-Craft2020}. Once the model experiences plasticity, changes in these material parameters are able to account for markedly different behaviours immediately post-yield, while also capturing the tendency of the systems to evolve towards similar states, whereas the previously developed model features trajectories that vary massively as the parameter values are changed, even though the initial state variables are the same. This again highlights the superior physical accuracy of the model developed in this paper, as it respects the observed tendency of soils and rocks to converge to similar stress ratios under continued loading, even if the initial behaviour is substantially different \citep{Bandini2011,Lade2005}.\\\\
While in this paper we have focused on the modelling of a large-grained and porous sandstone, the formalism is suitable for other types of minerals and can be applied to any sort of rock featuring distinct and crushable grains such as limestones \citep{Abdallah2021}.\\\\
We compare the predictions of the LSA against the results of the FEM analyses with respect to the initial band thickness in \figref{fig:LSA_vs_FEM_comparison}, where we have conducted additional simulations with $\chi_{0}\in\{0.25,0.75\}$ and $p_{0}\in\{150,250\}$ to gain a clearer understanding of the variation of the system with the problem parameters.
We observe in \figref{fig:LSA_vs_FEM_comparison} that the LSA predictions generally correspond very well to the width of the shear band in the FEM simulations inferred by the technique in Appendix~\ref{sec:fem_localisation_width}. The only visible exception is the width inferred by fitting over $B$ when $\chi_{0}$ is equal to zero, where the fitting underestimates the width of the band due to an influence from the initial imperfection, which exerts a stronger influence in this case where the values of $B$ the fitting is conducted over are very small. Changing the initial solid fraction of the system $\chi_{0}$ does not lead to substantially different initial shear band widths, in either constant volume or constant confining stress systems. We observe that the finite element fittings with $\gamma_{21}^{p}$ also demonstrate the same nonmonotonicity with $\chi_{0}$ as the results from the linear stability analyses that is observed in both load conditions. By contrast, changing the initial confining pressure can have a dramatic effect, with the shear band width at $p_{0}=100$~MPa being much wider than the width observed at $p_{0}=300$~MPa for both loading conditions. However, the change between the higher confining stresses is relatively small. Both the LSA results and FEM fittings capture these trends, and there is very close agreement in the predicted widths obtained with each method.\\\\
In the constant volume load case the LSA predicted gradual delocalisation while the inferred shear band widths from the FEM analyses remain stable. By contrast, the post-localisation behaviour under constant confining stress in the FEM simulations corresponds very well with the predictions of the LSA. The shear band rapidly delocalises, as the LSA predicted, and the speed of this delocalisation depends strongly on the values of $\chi_{0}$ and $p_{0}$, as predicted. The only meaningful difference between the LSA predictions and the results of the FEM simulations are for the $\chi_{0}=1.0$ and $p_{0}=100.0$~MPa systems, which delocalise more slowly than the LSA predicts, and the interaction with the boundaries of the system that occurs for the most rapidly delocalising systems. We may observe that the underlying reason for the interaction with the boundaries is that the shear bands under this loading condition concentrate their strain rate at the boundaries of the band, which march steadily outwards. This behaviour is illustrated in \figref{fig:illustrative_spatial_distribution_constant_confining_stress}.
\figref{fig:illustrative_spatial_distribution_constant_confining_stress} demonstrates clearly that while the shear band continues growing in the centre, the shearing occurs preferentially at the edges of the band, causing it to grow outwards until the boundary conditions interfere with further propagation.
\begin{figure}[H]
	\centering
	\includegraphics{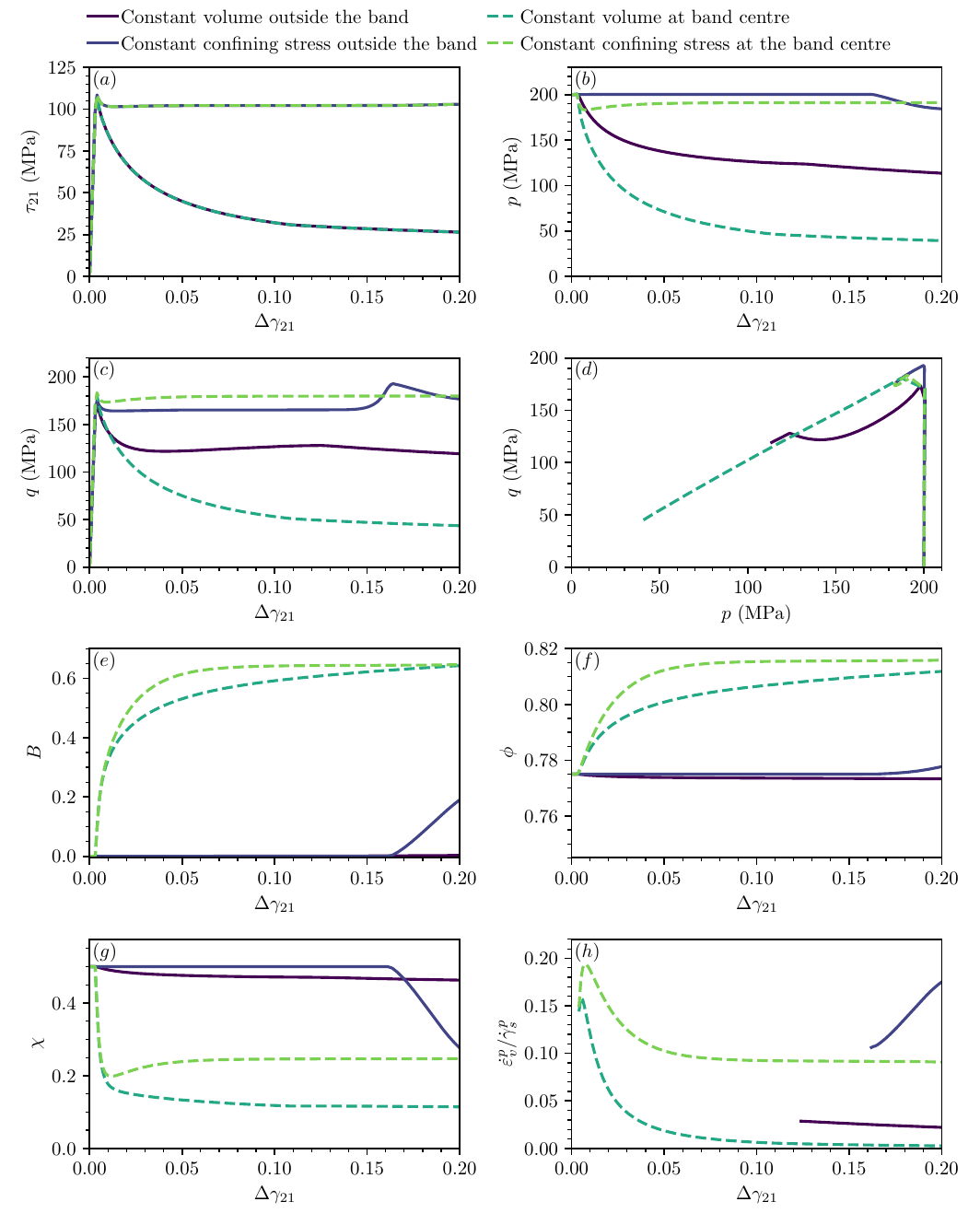}
	\caption{The results of simulations of shearing at constant volume and constant confining stress with $\chi_{0}=0.5$ and an initial confining stress of 200~MPa, considering points either near the edge of the system or at the centre of the band. (a) The shear stress $\tau_{21}$ against the increment of (homogeneous) shear strain $\Delta\gamma_{21}$, (b) the mean stress $p$ against the increment of (homogeneous) shear strain $\Delta\gamma_{21}$, (c) the deviatoric stress invariant $q$ against the increment of (homogeneous) shear strain $\Delta\gamma_{21}$, (d) the deviatoric stress invariant $q$ against the mean stress $p$, (e) the breakage index $B$ against the increment of shear strain $\Delta\gamma_{21}$, (f) the solid fraction $\phi$ against the increment of (homogeneous) shear strain $\Delta\gamma_{21}$, (g) the relative solid fraction $\chi$ against the increment of (homogeneous) shear strain $\Delta\gamma_{21}$, and (h) the ratio of the plastic volumetric strain rate and the plastic shear strain rate invariant against the increment of (homogeneous) shear strain $\Delta\gamma_{21}$.}
	\label{fig:FEM_material_comparison}
\end{figure}
In \figref{fig:FEM_material_comparison} we examine the different evolutions of the state variables inside and outside the shear band, taking the central values of $\chi_{0}=0.5$ and $p_{0}=200$~MPa as representative of the general behaviours. The values at the centre of the shear band are those for the elements with the highest value of $B$ at the end of the simulation, while the values outside the band are taken to be the values five elements in from the top of the system, to avoid distortions from the boundary. As expected, $\tau_{12}$ matches perfectly due to the need to maintain stress equilibrium, while $p$ and $q$ show similar tendencies inside and outside the band, but the specific values vary due to the different evolution of the $\tau_{22}$ and $\tau_{33}$ stresses as $B$ and $\phi$ evolve. Where the most dramatic differences are observed is in the evolution of the breakage and solid fraction. Here, the local concentration of strain (noting that the amount of strain experienced in this central element is much higher than the global homogeneous strain) drives the breakage higher rapidly. This drives in turn a substantial increase in the solid fraction, both in order to stay within the bounds imposed by \eqref{eq:phi_min} and \eqref{eq:phi_max}, but also to maintain the necessary stress equilibrium, keeping in mind that for a fixed elastic strain, an increase in $B$ will reduce the stress while an increase in $\phi$ will increase the stress (\textit{via} changing $\rho$). We also observe that the constant confining stress FEM simulation finds steady-state values of $B$ and $\phi$ reasonably early in the simulation, and correspondingly maintains an essentially constant $\chi$ in the band after an initial rapid drop. Similarly, this system maintains a relatively constant tendency towards plastic compaction. It is only as the band delocalises that we observe any substantial evolution of these quantities at the point that was (initially) outside the band. The constant volume FEM simulation grows the breakage and solid fraction values slightly more slowly inside the band, but by the end of the simulation the values are almost equal to the constant confining stress system. Where it differs significantly is in a more-or-less continuous decrease (at a decreasing rate) in the values of $\chi$ and $\dot{\varepsilon}_{v}^{p}/\dot{\gamma}_{s}^{p}$, suggesting the system attains a state of almost perfectly isochoric plastic shearing in the band. Outside the band, we observe essentially no change in these parameters, other than a very slight decrease in $\chi$ due to elastic dilation. At a certain point the outside of the band enters plasticity and is slightly compactive, but the evolution of the state variables is minimal.\\\\
One of the key differences between the single element tests and the finite element systems is that the finite element is able to account for structural features that vary in space. The most important spatial variations (other than the localisation into the shear band itself) are the outwards spreading of the shear strain rate observed for the constant confining stress simulations, and the dilation observed for the constant volume tests. These two phenomena cause the band to either experience less shear straining at the centre than would be expected, or permit additional compaction to be accommodated in the band than the boundary conditions allow for single element tests. Beyond the purely mechanical behaviour, the spatially complex evolution of the state variables also influence the permeability of the system to fluids. We can calculate the ratio of the current permeability at a point to its original permeability using \eqref{eq:permeability_ratio}.
\begin{figure}[H]
	\centering
	\includegraphics{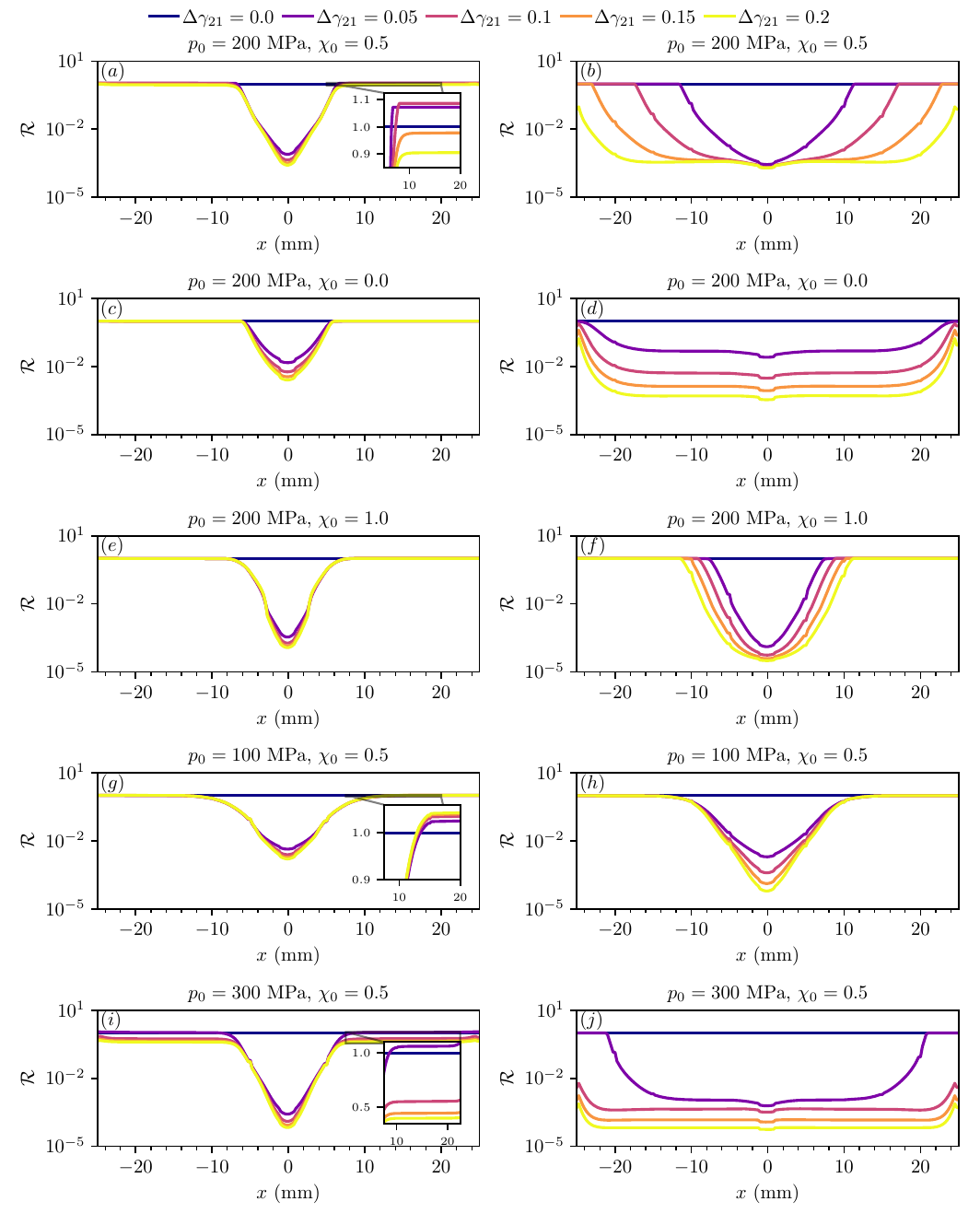}
	\caption{The ratio of the current permeability to the original permeability, for a selection of finite element system simulated, at a selection of key strain increments. The first column of subfigures ((a), (c), (e), (g), (i)) are the simulations under constant volume, while the second column of subfigures ((b), (d), (f), (h), (j)) are the simulations under constant confining stress. The insets in subfigures (a) and (i) show the permeability ratio in the region defined by the $x$-coordinate 5 to 20, with ratios less than 0.9 not shown.}
	\label{fig:permeability_cross_section}
\end{figure}
\figref{fig:permeability_cross_section} demonstrates the very large differences in permeability evolution between the simulations under constant volume and under constant confining stress. The general trend of the constant confining stress simulations is to have a relatively homogeneous decrease in the permeability ratio across the system, other than where the effect of the boundaries becomes significant. The exceptions to this are the systems that are slowest to delocalise, those with $\chi_{0}=1.0$ and $p_{0}=100$~MPa. Due to the greater concentration of shearing at the centre of the band thanks to the absence of complete delocalisation, these systems induce more grain crushing and porosity reduction at their centres, causing in turn large decreases in the permeability. Turning to the systems sheared under constant volume, we observe that in general the systems that favour grain breakage over pore collapse (higher initial values of $\chi_{0}$) are those that feature the greatest reductions in permeability ratio, indicating that within the proposed model the increase in tortuosity induced by the presence of large numbers of small grains plays a more important role than the reduction in pore space when it comes to reducing the permeability ratio. Of particular interest for the constant volume loading case is the changes induced outside the shear band. For the systems with $\chi_{0}=0.5$ and $p_{0}=200$~MPa and $p_{0}=300$~MPa, we observe changes that depend on the amount of shearing applied to the system. For low to moderate values of homogeneous shear straining, we see that the dilation that occurs outside the band leads to increases in the permeability ratio, facilitating fluid flow. However, after a certain point small amounts of breakage are induced which reverse this trend, and we see a small reduction in permeability outside the band. Finally, for the system with $p_{0}=100$~MPa, we observe that increasing the shear strain monotonically increases the permeability outside the band due to dilation.
\begin{figure}[H]
	\centering
	\includegraphics{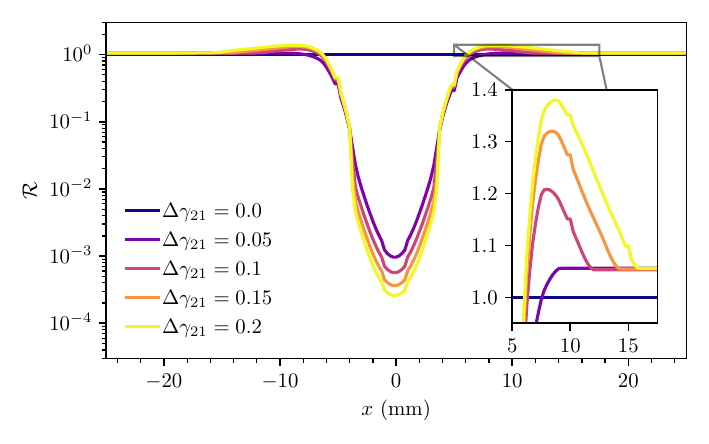}
	\caption{The ratio of the current permeability to the original permeability for the constant volume finite element system with $\zeta=1$, $\chi_{0}=0.5$ and $p_{0}=200$~MPa, at a selection of key strain increments. The inset shows the permeability ratio in the region defined by the $x$-coordinate 5 to 17.5, with ratios less than 0.95 not shown.}
	\label{fig:permeability_cross_section_zeta_1}
\end{figure}
In \figref{fig:permeability_cross_section_zeta_1}, we examine the permeability ratio of a system sheared under constant volume using the central parameter values, except that $\zeta=1.0$, meaning that the system favours dilation more than the other systems examined in this paper. We observe that in this case, we have a marked increase in the permeability of the system in two bands on either side of the principal shear band, with the permeability increasing more with increasing shearing. For all values of shear strain increment that these bands appear in, the maximal permeability ratio not only increases, but the bands become wider. These numerical results correspond to experimental observations on sands which have shown that even small amounts of grain breakage concentrated in shear bands grain substantially reduce the permeability \citep{Feia2016,Benammar2025}, as well as experimental observations on Fontainebleau sandstone, which have shown the presence of dilating zones with increased porosity outside compacting shear bands \citep{ElBied2002}, and that these dilating zones are capable of absorbing pore fluid expelled from the compacting region due to their much higher porosity and permeability \citep{Sulem2006}. In the context of faults, in which we can observe shearing much greater than 0.2, we can expect similar systems to demonstrate even greater increases in the permeability, up to the limit defined by the minimum solid fraction of the model.
\begin{figure}[H]
	\centering
	\includegraphics{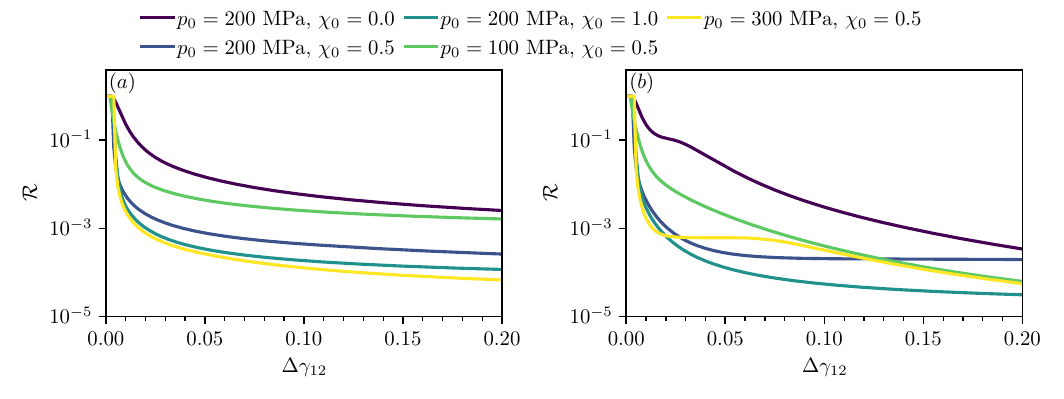}
	\caption{The permeability ratio at the centre of the band against the homogeneous shear strain for the systems sheared under (a) constant volume and (b) constant confining stress.}
	\label{fig:permeability_at_centre}
\end{figure}
In \figref{fig:permeability_at_centre} we may observe the evolution of the minimum value of the permeability ratio in the system against the homogeneous shear strain. For both systems we observe that changing the initial relative solid fraction $\chi_{0}$ and increasing the initial confining pressure $p_{0}$ favour greater reductions in the permeability ratio. The constant volume systems demonstrate a relatively constant shape, with fast initial reductions in permeability ratio followed by much longer periods of smaller but steady decline. While two of the constant confining stress systems show this same pattern (notably, the two systems that do not delocalise during the simulation), the three other systems either reach an essentially steady state in the value of the permeability ratio ($\chi_{0}=0.5$), or show ``kinks'' as the rate of permeability reduction decreases substantially before beginning to decline at steady rates. This behaviour is linked to the structural evolution of the band as it delocalises, with plateaus in the permeability ratio being linked to the shearing (and hence evolution of $B$ and $\phi$) largely occurring at the edges of the band, rather than in the centre. For the two systems where the band reaches the edge of the system, the evolution proceeds quasi-homogeneously, with the only variations due to the boundaries and the initial imperfection. Hence, the centre resumes its evolution. As the $\chi_{0}=0.5$ system only just completes its delocalisation over the course of the simulation, the system is still evolving towards the quasi-homogeneous evolution by accumulating breakage and porosity reduction at the boundaries, and so no meaningful evolution of the permeability ratio at the centre occurs. We note that under constant confining stress, after the $p_{0}=300$~MPa system has completed its delocalisation, there is very little difference between its permeabiltiy ratio and that of the $p_{0}=100$~MPa system, demonstrating that the continued concentration of shearing at the centre of the band for the latter system is able to drive similar amounts of grain crushing and pore collapse as the greater confining stress induces with more diffused shearing.\\\\
For constant confining stress system, it is the $\chi_{0}=1.0$ system that has the greatest amount of permeability reduction at the end of the simulation, while it has the second greatest reduction under constant volume, indicating that not only are the densest (least porous) systems the ones that are likely to have the lowest initial permeability, but that they are also likely to experience the greatest permeability reduction during shear band formation. Experimental values for the permeability of Bentheim sandstone are typically in the range of 1--7.5$\times10^{-13}$~m$^{2}$ \citep{Vajdova2004,Fazio2023}, so applying our reductions in permeability ratio, we could expect to see permeability values at the centre of the shear band ranging from 3$\times10^{-18}$ to 1.75$\times10^{-15}$, depending on the initial relative solid fraction and the loading conditions applied to the system. The lower part of this permeability range is more typical of claystones than sandstones \citep{Neuzil2019}, highlighting the dramatic effect of breakage evolution and pore collapse on the hydraulic properties of the material, and further underlining the importance of mechanical models that can account for this effect in the accurate modelling of faults, as such dramatic permeability changes will substantially influence the hydraulic and thermal coupling that governs the behaviour of the system. As the model developed in \citet{Collins-Craft2020} has no representation of the porosity (and plastic volumetric compaction can occur indefinitely in that model, without regard to physical limits), it is unable to represent the permeability in any meaningful way, given the strong dependence on the pore volume of the material.\\\\
In single-point simulations, we observe that the new model generally exhibits slower changes to its underlying state variables, in particular the breakage index $B$. This occurs as the new model also includes the solid fraction, and as the two variables co-evolve the system typically finds states that lead to $\omega$ increasing, and hence the rate of breakage decreasing. By contrast, in the original model, the rate of breakage is meaningfully modified only by the value of $B$, so it is only as $B\rightarrow1$ that the rate of breakage slows down. As a consequence of this behaviour, the new model offers physically more realistic evolutions of the GSD with increasing strain. Changes in the initial relative solid fraction modify the initial behaviour of the system in a way similar to changing $\bar{\omega}$ in the original model, but the balancing of $B$ and $\phi$ evolution in the new model causes the systems to tend towards similar states, regardless of the starting position, while the original model can demonstrate extremely divergent behaviour depending on the value of $\bar{\omega}$. We also observe that our new model consistently predicts the appearance of shear bands earlier than the original model, specifically at the initiation of plasticity. While in real geomaterials it is generally not possible to cleanly distinguish when plasticity commences, at the very least we can note that the early localisation of the new model is much closer to observed experimental phenomena in triaxial tests on sand \citep{Desrues2018}, in which shear bands appear well before the peak stress. Our model could easily be adapted to no longer cleanly separate between elastic and plastic behaviours by reframing it in $h^{2}$-plasticity \citep{Einav2012}, although we have elected to remain in a more classical plasticity framework in this paper for reasons of numerical efficiency.
\section{Conclusion}\label{sec:conclusion}
In this paper, we developed a constitutive model in the Cosserat Breakage Mechanics framework introduced in \citet{Collins-Craft2020}. This model features a number of improvements that increase its physical fidelity, namely density-dependent elasticity and shear strength, the presence of cohesive strength and the inclusion of the solid fraction $\phi$ as a state variable, enable the modelling of dilation at lower confining stresses, and the coevolution of the grain size distribution with the pore space at higher confining stresses.\\\\
This model was then calibrated against experimental data obtained by other authors on Bentheim sandstone, demonstrating the importance of the new model features to being able to accurately calibrate the yield surface to data obtained over the full range of stresses up to to the crushing pressure. We then conducted sensitivity analyses of the new model and comparisons with the model presented in \citet{Collins-Craft2020} to examine the mechanical behaviour as the initial solid fraction $\phi_{0}$ and the initial confining stress $p_{0}$ were varied. These mechanical analyses were accompanied by predictions of the initial localisation behaviour of the model, determined by linear stability analyses. These analyses highlight that localisation of the new model occurs directly upon the start of plasticity and that the resultant shear bands are compactive at confining stresses typical of earthquake nucleation. In the case of constant confining stress, these bands further demonstrate a tendency to delocalise.\\\\
The localisation results obtained by linear stability analyses were subsequently confirmed by one-dimensional finite element simulations that showed close matching for the initial localisation width, and in the case of constant confining stress, confirmed the predicted delocalisation behaviour. Under constant volume conditions, similar to the conditions faults experience during fast undrained shearing, certain systems demonstrate not only the presence of compacting shear bands, but regions of dilation immediately outside the band. Using a modified Kozeny--Carman permeability law, we are able to analyse the evolution of the permeability across the system and observe both orders-of-magnitude reduction of the permeability within the shear band, as well as increases in the permeability in the dilating regions outside the band, corresponding to experimental observations made on similar sandstones.\\\\
This work has substantially increased the physical fidelity of the Cosserat Breakage Mechanics model family and broadened the range of materials that it is able to accurately model. The inclusion of an additional state variable gives the model the capability to model phenomena that were previously inaccessible, and enables a coupling of the model with hydrological quantities such as pore fluid flow. This work can thus serve as the mechanical component of coupled thermo-hydro-(chemo)-mechanical models that can fully elucidate the formation of the shear bands that make up the core of seismogenic faults.
\section*{Acknowledgements}\label{sec:acknowledgements}
The first author acknowledges the support of the Marie Sk\l odowska-Curie Actions program under the Horizon Europe research and innovation framework program (Grant agreement ID 101064805 LEMMA). The second author acknowledges the European Research Council’s (ERC) support under the European Union’s Horizon 2020 research and innovation programme (Grant agreement no. 101087771 INJECT). Views and opinions expressed are however those of the authors only and do not necessarily reflect those of the European Union or Horizon Europe. Neither the European Union nor the granting authority can be held responsible for them.
\section*{Open Research}
The code required to run the simulations described in this paper is available in a GitHub repository that has been archived on \href{https://archive.softwareheritage.org/swh:1:dir:14be36bbbc9d1cc6bf869a2649c7f7d1a794a229;origin=https://github.com/nickcollins-craft/The-influence-of-grain-crushing-and-pore-collapse-on-the-formation-of-faults;visit=swh:1:snp:dfd9fd7194df234314df63e7aabb8937f98dfa18;anchor=swh:1:rev:0ce2d9e3fad6da445bba1bc209b472ca775cd604}{Software Heritage}, or alternatively as a deposit on \href{https://doi.org/10.5281/zenodo.17199356}{Zenodo}. The data outputs of the codes are also available on \href{https://doi.org/10.5281/zenodo.17199464}{Zenodo}.
\appendix
\section{Model derivation}\label{sec:model_derivation}
\subsection{Thermodynamic admissibility}\label{sec:thermodynamic_admissibility}
We wish to show that the model specified in \secref{sec:constitutive_model} is thermodynamically admissible (that is, both the internal energy and the dissipation are never negative).\\\\
Firstly, we consider the internal energy of the system. Our definitions of $B$, $\theta_{\gamma}$ and $\theta_{\kappa}$ ensure that $B\in[0,1]$, $\theta_{\gamma}\in[0,\frac{2}{5-\alpha}]$ and $\theta_{\kappa}\in[0,\frac{4}{7-\alpha}]$ (all positive for typical values of $\alpha$ of $2.5\thicksim2.7$), the maximum grain size $d_{\textrm{Max}}>0$, the density $\rho>0$ (as is the unstressed solid density $\rho_{s}^{\star}$), and the tensors $C_{ijkl}^{e}$ and $D_{ijkl}^{e}$, when transformed to two-dimensional matrices, are positive definite (provided physically admissible values of $\bar{K}$ and $\bar{G}$ are chosen), meaning their products with $\gamma_{ij}^{e}$ and $\kappa_{ij}^{e}$ (when both transformed to one-dimensional vectors) will both be positive. As such, by inspection we can conclude that \eqref{eq:u_function_with_rho_star} is always non-negative.\\\\
Demonstrating the positiveness of the dissipation is substantially more elaborate. We start by declaring the Clausius--Duhem inequality for an isothermal process in the Cosserat continuum:
\begin{equation}
	\mathcal{D}=\tau_{ij}\dot{\gamma}_{ij}+\mu_{ij}\dot{\kappa}_{ij}-\left(\dot{\mathcal{U}}-\mathcal{U}\frac{\dot{\rho}}{\rho}\right)\geq0,\label{eq:clausius_duhem}
\end{equation}
where $\mathcal{U}$ is the internal energy (we add the hat in \eqref{eq:u_function_with_rho_star} to distinguish between the \textit{value} of internal energy (without hat) and the \textit{function} that returns its value given the state variables (with hat)) and $\mathcal{D}$ is the dissipation. From \eqref{eq:u_function_with_rho_star} we recover the time rate of the internal energy:
\begin{equation}
	\dot{\mathcal{U}}=\frac{\partial\hat{\mathcal{U}}}{\partial\gamma_{ij}^{e}}\dot{\gamma}_{ij}^{e}+\frac{\partial\hat{\mathcal{U}}}{\partial\kappa_{ij}^{e}}\dot{\kappa}_{ij}^{e}+\frac{\partial\hat{\mathcal{U}}}{\partial\rho}\dot{\rho}+\frac{\partial\hat{\mathcal{U}}}{\partial B}\dot{B}.\label{eq:internal_energy_rate}
\end{equation}
The mass balance equation is given by
\begin{equation}
	\dot{\rho}+\rho\dot{u}_{i,i}=0.\label{eq:mass_balance}
\end{equation}
This can equivalently be written as
\begin{equation}
	\frac{\dot{\rho}}{\rho}=\dot{\gamma}_{ij}\delta_{ij}=\dot{\varepsilon}_{v}.\label{eq:mass_balance_in_terms_of_strains}
\end{equation}
Now, substituting \eqref{eq:internal_energy_rate} and \eqref{eq:mass_balance_in_terms_of_strains} back into \eqref{eq:clausius_duhem}, we obtain:
\begin{equation}
	\mathcal{D}=\tau_{ij}\dot{\gamma}_{ij}+\mu_{ij}\dot{\kappa}_{ij}-\left(\frac{\partial\hat{\mathcal{U}}}{\partial\gamma_{ij}^{e}}\dot{\gamma}_{ij}^{e}+\frac{\partial\hat{\mathcal{U}}}{\partial\kappa_{ij}^{e}}\dot{\kappa}_{ij}^{e}+\frac{\partial\hat{\mathcal{U}}}{\partial\rho}\dot{\rho}+\frac{\partial\hat{\mathcal{U}}}{\partial B}\dot{B}-\mathcal{U}\dot{\gamma}_{ij}\delta_{ij}\right)\geq0,\label{eq:clausius_duhem_first_substitution}
\end{equation}
Then, using the fact that $\dot{\rho}=\rho\dot{\gamma}_{ij}\delta_{ij}$, and substituting \eqref{eq:strain_decomposition} and \eqref{eq:curvature_decomposition} into \eqref{eq:clausius_duhem_first_substitution}, we get
\begin{align}
	\mathcal{D}&=\tau_{ij}(\dot{\gamma}_{ij}^{e}+\dot{\gamma}_{ij}^{p})+\mu_{ij}(\dot{\kappa}_{ij}^{e}+\dot{\kappa}_{ij}^{p})-\left(\frac{\partial\hat{\mathcal{U}}}{\partial\gamma_{ij}^{e}}\dot{\gamma}_{ij}^{e}+\frac{\partial\hat{\mathcal{U}}}{\partial\kappa_{ij}^{e}}\dot{\kappa}_{ij}^{e}+\frac{\partial\hat{\mathcal{U}}}{\partial\rho}\rho(\dot{\gamma}_{ij}^{e}+\dot{\gamma}_{ij}^{p})\delta_{ij}+\frac{\partial\hat{\mathcal{U}}}{\partial B}\dot{B}-\mathcal{U}(\dot{\gamma}_{ij}^{e}+\dot{\gamma}_{ij}^{p})\delta_{ij}\right)\geq0,\nonumber\\
	&=\left(\tau_{ij}-\frac{\partial\hat{\mathcal{U}}}{\partial\gamma_{ij}^{e}}-\left(\rho\frac{\partial\hat{\mathcal{U}}}{\partial\rho}-\mathcal{U}\right)\delta_{ij}\right)\dot{\gamma}_{ij}^{e}+\left(\mu_{ij}-\frac{\partial\hat{\mathcal{U}}}{\partial\kappa_{ij}^{e}}\right)\dot{\kappa}_{ij}^{e}+\left(\tau_{ij}-\left(\rho\frac{\partial\hat{\mathcal{U}}}{\partial\rho}-\mathcal{U}\right)\delta_{ij}\right)\dot{\gamma}_{ij}^{p}\nonumber\\
	&\quad\quad+\mu_{ij}\dot{\kappa}_{ij}^{p}-\frac{\partial\hat{\mathcal{U}}}{\partial B}\dot{B}\geq0.\label{eq:clauius_duhem_fully_subbed}
\end{align}
\eqref{eq:clauius_duhem_fully_subbed} must hold true for arbitrary values of $\dot{\gamma}_{ij}^{e}$, $\dot{\kappa}_{ij}^{e}$, $\dot{\gamma}_{ij}^{p}$, $\dot{\kappa}_{ij}^{p}$ and $\dot{B}$. As purely elastic processes with $\dot{\gamma}_{ij}^{p}$, $\dot{\kappa}_{ij}^{p}$ and $\dot{B}$ all equal to zero must not dissipate energy, we conclude the terms inside the brackets multiplying $\dot{\gamma}_{ij}^{e}$ and $\dot{\kappa}_{ij}^{e}$ must be identically zero so that the relation always holds. We hence define
\begin{align}
	\tau_{ij}^{e}&=\frac{\partial\hat{\mathcal{U}}}{\partial\gamma_{ij}^{e}},\label{eq:elastic_stress}\\
	\eta&=\frac{\partial\hat{\mathcal{U}}}{\partial\rho},\label{eq:chemical_potential}\\
	p^{t}&=\rho\eta-\mathcal{U},\label{eq:thermodynamic_pressure}\\
	\tau_{ij}&=\tau_{ij}^{e}+p^{t}\delta_{ij},\label{eq:true_stress}\\
	\mu_{ij}&=\frac{\partial\hat{\mathcal{U}}}{\partial\kappa_{ij}^{e}},\label{eq:couple_stress}\\
	E_{B}&=-\frac{\partial\hat{\mathcal{U}}}{\partial B},\label{eq:breakage_energy}
\end{align}
where we will call $\tau_{ij}^{e}$ the elastic stress, $\eta$ is the chemical potential, $p^{t}$ is a thermodynamic pressure, and $E_{B}$ is the breakage energy. The couple stresses and breakage energy are standard for Cosserat Breakage Mechanics, but the more refined treatment of the thermodynamics demands consideration of the chemical potential and thermodynamic pressure to correctly obtain the stresses. However, we note that in typical experimental conditions for dry granular media the thermodynamic pressure $p^{t}$ is negligible. Then, the dissipation is given by what remains after removing the parts acting on the elastic rates:
\begin{equation}
	\mathcal{D}=\tau_{ij}^{e}\dot{\gamma}_{ij}^{p}+\mu_{ij}\dot{\kappa}_{ij}^{p}+E_{B}\dot{B}\geq0.\label{eq:dissipation_relation}
\end{equation}
Now, we need to obtain a time-evolution law for the solid fraction. Re-arranging \eqref{eq:phi_def} and applying the time derivative, we get
\begin{equation}
	\dot{\rho}=\dot{\phi}\rho_{s}+\phi\dot{\rho}_{s},\label{eq:rho_dot_with_varying_rho_s}
\end{equation}
Then, after using our constitutive assumption \eqref{eq:dot_rho_s}, we have the total density rate
\begin{equation}
	\dot{\rho}=\rho_{s}\dot{\phi}+\phi\rho_{s}\chi\dot{\varepsilon}_{v}^{e},\label{eq:total_density_rate}
\end{equation}
which can be equated with the time rate of the density given by rearranging \eqref{eq:mass_balance_in_terms_of_strains} and substituting in \eqref{eq:phi_def}:
\begin{align}
	\rho_{s}\dot{\phi}+\phi\rho_{s}\chi\dot{\varepsilon}_{v}^{e}&=\phi\rho_{s}\dot{\varepsilon}_{v},\nonumber\\
	\dot{\phi}&=\phi\dot{\varepsilon}_{v}-\phi\chi\dot{\varepsilon}_{v}^{e},\nonumber\\
	&=\phi(\dot{\varepsilon}_{v}^{e}+\dot{\varepsilon}_{v}^{p})-\phi\chi\dot{\varepsilon}_{v}^{e},\nonumber\\
	&=\phi\left[(1-\chi)\dot{\varepsilon}_{v}^{e}+\dot{\varepsilon}_{v}^{p}\right],\label{eq:dot_phi}
\end{align}
which recovers the expression given in \citet{Alaei2021}. This also allows us to attribute elastic and plastic components to the rate of the solid fraction:
\begin{align}
	\dot{\phi}^{e}&=\phi(1-\chi)\dot{\varepsilon}_{v}^{e},\label{eq:dot_phi_e}\\
	\dot{\phi}^{p}&=\phi\dot{\varepsilon}_{v}^{p}.\label{eq:phi_dot_p}
\end{align}
Now, we return to the dissipation expression given in \eqref{eq:dissipation_relation} and rearrange it
\begin{align}
	d&=\tau_{ij}^{e}\dot{\gamma}_{ij}^{p}+\mu_{ij}\dot{\kappa}_{ij}^{p}+E_{B}\dot{B}\geq0,\nonumber\\
	&=\left(p^{e}\delta_{ij}+s_{ij}^{e}\right)\left(\frac{1}{3}\dot{\varepsilon}_{v}^{p}\delta_{ij}+\dot{e}_{ij}^{p}\right)+\left(\frac{1}{3}\mu_{kk}\delta_{ij}+m_{ij}\right)\left(\frac{1}{3}\dot{\kappa}_{kk}^{p}+\dot{z}_{ij}^{p}\right)+E_{B}\dot{B}\geq0,\nonumber\\
	&=p^{e}\dot{\varepsilon}_{v}^{p}+s_{ij}^{e}\dot{e}_{ij}^{p}+m_{ij}\dot{z}_{ij}^{p}+E_{B}\dot{B}\geq0,\nonumber\\
	&=\frac{p^{e}}{\phi}\dot{\phi}^{p}+s_{ij}^{e}\dot{e}_{ij}^{p}+m_{ij}\dot{z}_{ij}^{p}+E_{B}\dot{B}\geq0,\nonumber\\
	&=\frac{p^{e}}{\phi}\dot{\phi}^{p}+q\dot{\gamma}_{s}^{p}+E_{B}\dot{B}\geq0,\label{eq:dissipation_in_terms_of_invariants}
\end{align}
where to pass from the first line to the second line we expanded the stress and couple-stress terms into their trace and deviatoric parts, to pass from the second line to the third we applied our assumption that the trace parts of the couple-stress do no work (nor store any energy), from the third to the fourth we used \eqref{eq:phi_dot_p}, and from the fourth to the fifth we used the fact that the weighting factors in our invariants are those for which the equality $q\dot{\gamma}_{s}^{p}=s_{ij}\dot{e}_{ij}^{p}+m_{ij}\dot{z}_{ij}^{p}$ holds \citep{Collins-Craft2019}. Now, substituting in the flow rules \eqref{eq:B_dot}, \eqref{eq:phi_p_dot} and \eqref{eq:gamma_s_p_dot} we obtain:
\begin{align}
	0\leq&\frac{p^{e}}{\phi}\cdot\lambda\left(\sqrt{\frac{E_{B}}{E_{c}}}(1-B)-\zeta\chi\right)\frac{2\phi(1-B)}{p^{e}}\sqrt{\frac{E_{B}}{E_{c}}}\sin^{2}(\omega)+\lambda\frac{2{q}^{2}}{(Mp^{e}+\phi(1-B)c)^{2}}\nonumber\\
	&\quad+E_{B}\lambda\left\langle\sqrt{\frac{E_{B}}{E_{c}}}(1-B)-\zeta\chi\right\rangle\frac{2(1-B)}{\sqrt{E_{B}E_{c}}}\cos^{2}(\omega).\label{eq:dissipation_equation}
\end{align}
We have that the parameters $E_{c},\zeta,c$ are all positive, the variables $\phi,B,E_{B}$ likewise, the function $M_{0}$ is positive for all admissible values which in turn guarantees that $M$ is also positive, $0\leq\omega\leq\pi/2$ and $\lambda$ is non-negative. We consider only the case of positive $p^{e}$ in this model, as tensile failure is governed by different physics (macroscopic cracking) that requires dramatically different models to address. In combination with the Macaulay brackets all these restrictions on the values of the variables guarantee that the third term is non-negative. The second term is composed of a ratio of two squared terms and thus is straightforwardly never negative. In the case of $F\geq0$, the first term is also straightforwardly non-negative given our restrictions on the variable values. For $F<0$, the analysis is slightly more complicated. We return to \eqref{eq:y_mix} and rearrange it in terms of ${q}^{2}$ and $F$ to obtain ${q}^{2}=(1-F^{2})(Mp^{e}+\phi(1-B)c)^{2}$. Substituting this back in to \eqref{eq:dissipation_equation}, cancelling out the common factors and suppressing the third term as we have assumed $F<0$, we have
\begin{equation}
	(1-B)\sqrt{\frac{E_{B}}{E_{c}}}\sin^{2}(\omega)F+1-F^{2}\geq0.\label{eq:rearranged_dissipation_equation}
\end{equation}
As we have assumed $F<0$ we also have $\omega=\pi/2$ and hence $\sin^{2}(\omega)=1$, so we can remove it from the equation. We recall also that the value of $\sqrt{E_{B}/E_{c}}(1-B)$ is conditioned by our choice of $F$, namely $\sqrt{E_{B}/E_{c}}(1-B)=F+\zeta\chi$. Substituting this relationship back in to \eqref{eq:rearranged_dissipation_equation} we have
\begin{align}
	(F+\zeta\chi)F+1-F^{2}&\geq0,\nonumber\\
	F^{2}+\zeta\chi F+1-F^{2}&\geq0,\nonumber\\
	1+\zeta\chi F&\geq0.\label{eq:final_dissipation_inequality}
\end{align}
We see that this inequality is respected for all admissible values ($\zeta\in[0,1],\chi\in[0,1],F\in[-1,0)$) coherent with our assumption of $F<0$. This guarantees thermodynamic admissibility for $-1\leq F<0$, and in combination with the demonstrations above for $F\geq0$ we have demonstrated that the dissipation relationship \eqref{eq:dissipation_equation} is respected for \textit{all} admissible values of the variables.
\subsection{Plastic multiplier}\label{sec:plastic_multiplier}
In order to (analytically) close the model, the plastic multiplier $\lambda$ must be calculated. This is achieved by enforcing the consistency condition \textit{i.e.} $y=\dot{y}=0$ for continued inelastic loading. We exploit automatic differentiation to obtain the quantities $\frac{\partial y}{\partial\gamma_{kl}^{e}},\frac{\partial y}{\partial\kappa_{kl}^{e}},\frac{\partial y}{\partial\rho},\frac{\partial y}{\partial\phi},\frac{\partial y}{\partial B}$, which propagates the derivatives through the system using the chain rule. Hence, we obtain in symbolic form
\begin{equation}
	\lambda=\frac{\left(\frac{\partial y}{\partial\gamma_{kl}^{e}}+\frac{\partial y}{\partial\rho}\frac{\partial\rho}{\partial\dot{\gamma}_{kl}}+\frac{\partial y}{\partial\phi}\frac{\partial\dot{\phi}^{e}}{\partial\dot{\varepsilon}_{v}^{e}}\frac{\partial\dot{\varepsilon}_{v}^{e}}{\partial\dot{\gamma}_{kl}^{e}}\right)\dot{\gamma}_{kl}+\frac{\partial y}{\partial\kappa_{kl}^{e}}\dot{\kappa}_{kl}}{\frac{\partial y}{\partial\gamma_{kl}^{e}}\overline{\gamma_{kl}^{p}}+\frac{\partial y}{\partial\kappa_{kl}^{e}}\overline{\kappa_{kl}^{p}}+\frac{\partial y}{\partial\phi}\left(\frac{\partial\dot{\phi}^{e}}{\partial\dot{\varepsilon}_{v}^{e}}\frac{\partial\dot{\varepsilon}_{v}^{e}}{\partial\dot{\gamma}_{kl}^{e}}\overline{\gamma_{kl}^{p}}-\overline{\phi^{p}}\right)-\frac{\partial y}{\partial B}\overline{B}},\label{eq:lambda_analytical}
\end{equation}
where by $\overline{B}$, $\overline{\phi^{p}}$, $\overline{\gamma_{kl}^{p}}$, and $\overline{\kappa_{kl}^{p}}$ we mean the non-$\lambda$ parts of the flow rules \eqref{eq:B_dot}, \eqref{eq:phi_p_dot}, \eqref{eq:gamma_p_dot} and \eqref{eq:kappa_p_dot}.
\subsection{Incremental constitutive response}\label{sec:incremental_constitutive_response}
While the model is closed by calculating $\lambda$, it is convenient to write an expression for the incremental constitutive response, taking into account the appropriate cross-couplings between stresses and couple-stresses with the strain and curvature rates. We make a linear decomposition of $\lambda$:
\begin{equation}
	\lambda=\lambda^{\gamma}_{kl}\dot{\gamma}_{kl}+\lambda^{\kappa}_{kl}\dot{\kappa}_{kl}. \label{eq:lambda_decomposition}
\end{equation}
By differentiating the stress and couple-stress with respect to time, using \eqref{eq:lambda_decomposition} and inserting the appropriate flow rules, we may write the incremental constitutive relationship under continued plastic loading as
\begin{equation}
	\begin{bmatrix}
		\dot{\tau}_{ij}\\\dot{\mu}_{ij}
	\end{bmatrix}=\begin{bmatrix}
		E_{ijkl}^{ep} & F_{ijkl}^{ep} \\
		K_{ijkl}^{ep} & M_{ijkl}^{ep}
	\end{bmatrix}\begin{bmatrix}
		\dot{\gamma}_{kl}\\\dot{\kappa}_{kl}
	\end{bmatrix},
	\label{eq:incremental_constitutive_partial_tensors}
\end{equation}
where the continuum incremental elastoplastic stiffness matrices are given by
\begin{align}
	E^{ep}_{ijkl}&=\frac{\partial\tau_{ij}}{\partial\gamma_{kl}^{e}}+\frac{\partial\tau_{ij}}{\partial\rho}\frac{\partial\dot{\rho}}{\partial\dot{\gamma}_{kl}}-\left(\frac{\partial\tau_{ij}}{\partial\gamma_{kl}^{e}}\overline{\gamma_{kl}^{p}}+\frac{\partial\tau_{ij}}{\partial\kappa_{kl}^{e}}\overline{\kappa_{kl}^{p}}-\frac{\partial\tau_{ij}}{\partial B}\overline{B}\right)\lambda_{kl}^{\gamma},\label{eq:Eep}\\
	F^{ep}_{ijkl}&=\frac{\partial\tau_{ij}}{\partial\kappa_{kl}^{e}}
	-\left(\frac{\partial\tau_{ij}}{\partial\gamma_{kl}^{e}}\overline{\gamma_{kl}^{p}}+\frac{\partial\tau_{ij}}{\partial\kappa_{kl}^{e}}\overline{\kappa_{kl}^{p}}-\frac{\partial\tau_{ij}}{\partial B}\overline{B}\right)\lambda_{kl}^{\kappa},\label{eq:Fep}\\	
	K^{ep}_{ijkl}&=\frac{\partial\mu_{ij}}{\partial\gamma_{kl}^{e}}+\frac{\partial\mu_{ij}}{\partial\rho}\frac{\partial\dot{\rho}}{\partial\dot{\gamma}_{kl}}-\left(\frac{\partial\mu_{ij}}{\partial\gamma_{kl}^{e}}\overline{\gamma_{kl}^{p}}+\frac{\partial\mu_{ij}}{\partial\kappa_{kl}^{e}}\overline{\kappa_{kl}^{p}}-\frac{\partial\mu_{ij}}{\partial B}\overline{B}\right)\lambda^{\gamma}_{kl},\label{eq:Kep}\\	
	M^{ep}_{ijkl}&=\frac{\partial\mu_{ij}}{\partial\kappa_{kl}^{e}}-\left(\frac{\partial\mu_{ij}}{\partial\gamma_{kl}^{e}}\overline{\gamma_{kl}^{p}}+\frac{\partial\mu_{ij}}{\partial\kappa_{kl}^{e}}\overline{\kappa_{kl}^{p}}-\frac{\partial\mu_{ij}}{\partial B}\overline{B}\right)\lambda^{\kappa}_{kl}.\label{eq:Mep}
\end{align}
We emphasise here that the stress that is differentiated with respect to the state variables is the total stress, inclusive of the thermodynamic pressure, and we once again exploit automatic differentiation to obtain the necessary quantities.
\subsection{Permeability evolution}\label{sec:permeability_evolution}
As in this work, we are interested in the effect of changes in the microstructure on the global properties of the fault, we use the microstructural information to observe the changes in the permeability of the fault material as it evolves. It is well-established in the literature that the permeability plays a key role in controlling the behaviour of the fault, due to the tight coupling between hydraulic and mechanical effects \citep{Rice2006,Rattez2017,Rattez2018a,Rattez2018b,Rattez2018c,Stathas2023}. In general, we expect that increases in the solid fraction will in turn decrease the permeability of the system (and \textit{vice versa}) in a nonlinear way, as not only will the amount of space available for the fluid to flow through change, but the tortuosity will as well. Similarly, we expect that as the grains break, the permeability should reduce due to the increase in tortuosity of the sample. Following \citet{Nguyen2009} (who in turn follow \citet{Matyka2008}) in developing a modified Kozeny--Carman equation, we suppose a tortuosity function that depends on the solid fraction $T(\phi)\propto(1-\phi)^{-5}$. Then, for an initial permeability $k_{0}$ associated with the initial solid fraction $\phi_{0}$, the ratio of the current permeability $k$ to the original permeability is given by
\begin{equation}
	\label{eq:permeability_ratio}
	\mathcal{R}=\frac{k}{k_{0}}=\frac{(1-\phi)^{3}{T_{0}}^{2}{d_{H}}^{2}}{(1-\phi_{0})^{3}T^{2}{d_{H0}}^{2}}.
\end{equation}
\section{Determining the localisation width in finite element simulations}\label{sec:fem_localisation_width}
\begin{figure}[H]
	\centering
	\includegraphics{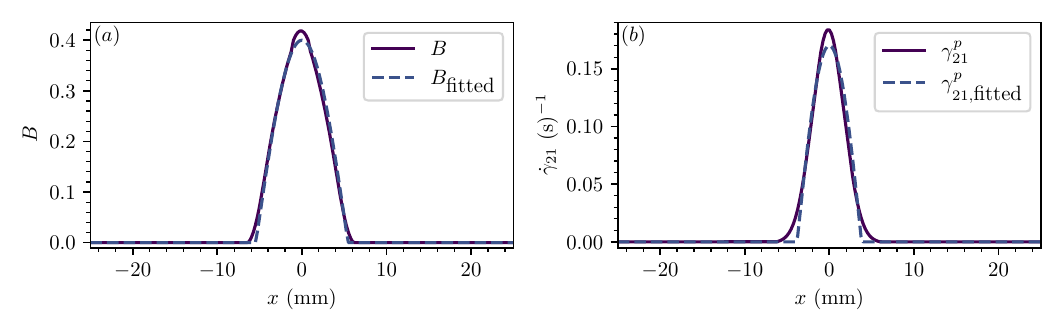}
	\caption{Representative best-fits using a cosine fitting function over the values generated by the simulation of constant volume shearing with $p_{0}=200$~MPa and $\chi_{0}=0.5$, taken at $\Delta\gamma_{21}=0.02$. (a) The distribution of generated and fitted values of $B$ over the spatial dimension $x$, and (b) the distribution of generated and fitted values of $\gamma_{21}^{p}$ over the spatial dimension $x$.}
	\label{fig:representative_cosine_fitting}
\end{figure}

While the finite element simulations return localisation widths that are visually clear, in order to obtain a quantitative measure, we must conduct a fitting process. In order to maintain coherence with the assumption of the linear stability analysis, we choose to fit a cosine curve to the localisation, fitting on the value of $B$ and the value of $\gamma_{21}^{p}$. We choose these two values as $\gamma_{21}^{p}$ is coherent with all plasticity models on which localisation analysis can be conducted, while $B$ gives an indication of the shear band width available only to models belonging to the Breakage Mechanics family. The best-fit cosine is re-found at each time step.\\\\
For a given time step, we first set $c$, a vertical shift parameter to be the value of $B$ or $|\dot{\gamma}_{21}|$ at the upper boundary of the system at that time step, so the function is able to match the values outside of the shear band. Then, we specify the fitting function:
\begin{equation}
	\hat{z}=\begin{cases}
		(A-c)\cos\left(\frac{2\pi x}{\Lambda}\right)+c & \text{if }\left|\frac{2\pi x}{\Lambda}\right|\leq\frac{\pi}{2},\\
		c & \text{if }\left|\frac{2\pi x}{\Lambda}\right|>\frac{\pi}{2},
	\end{cases}\label{eq:cosine_fitting_function}
\end{equation}
where $\hat{z}$ is our variable of interest, $A$ is the amplitude parameter and $\Lambda$ is the wavelength. This function fits a half-wavelength over the localisation, with the height controlled by $A$ and the width controlled by $\Lambda$.\\\\
At each time step, we then conduct an optimisation using NLopt \citep{NLopt}, in particular the COBYLA algorithm \citep{Powell1994,Powell1998}, to find the values of the parameters that minimise the least-squares difference between the values of the variable predicted by the fitting function and the values of the variables that were generated by the finite element simulation. We initialise the parameters with an initial guess of $\{A,\Lambda\}=\{\underset{x\in\mathcal{V}}{\argmax}(\hat{z},t),5.0\}$ where $t$ is the time at the current time step. We apply lower bounds of $\{A_{\textrm{lower}},\Lambda_{\textrm{lower}}\}=\{0.75\times\underset{x\in\mathcal{V}}{\argmax}(\hat{z},t),0.0001\}$ and upper bounds of $\{A_{\textrm{upper}},\Lambda_{\textrm{upper}}\}=\{1.5\times\underset{x\in\mathcal{V}}{\argmax}(\hat{z},t),100.0\}$. We only conduct the fitting if the any of the points are on the yield surface, as measured by $\underset{x\in\mathcal{V}}{\argmax}(y,t)\geq-1\times10^{-6}$, where $y$ is the value of the yield function. We show a representative comparison of the fitted curves in \figref{fig:representative_cosine_fitting}.
We can see that the shape of the localisation generated by the finite element simulation does not perfectly correspond to a cosine curve (nor is it perfectly centred within the simulation) but it is also not dramatically different and the cosine function satisfactorily captures the width of the shear band, which is the key requirement of the method. The quantitative value of the shear band width is obtained by dividing $\Lambda$ by two at each time step. In order to obtain the initial shear band width, we find the time step at which the fitting parameters are non-zero, and take the values from two time steps after that point, in order to allow the localisation to better reflect the material properties, rather than the initial imperfection.
\begin{refcontext}[sorting=nyt]
	\printbibliography
\end{refcontext}
\end{document}